\documentclass[twocolumn,letterpaper,amsmath,amssymb,floatfix,aps,superscriptaddress]{revtex4-1}

\usepackage{hyperref}
\usepackage{verbatim}
\usepackage{graphicx}
\usepackage{bm} 
\usepackage{ifpdf} 
\usepackage{dcolumn}  
\usepackage{epsfig}
\usepackage{color}

\newcommand{\Av}[1]{{\bf #1}}
\newcommand{\Avg}[1]{{\boldsymbol #1}}

\def\ln{{\operatorname{ln}}}
\def\det{{\operatorname{det}}}

\def\rmd{{\mathrm{d}}}
\def\rmi{{\mathrm{i}}}
\def\rme{{\mathrm{e}}}
\def\lB{\ell_{\mathrm{B}}}

\begin{document}

\title{Strong coupling electrostatics for randomly charged surfaces: Antifragility and effective interactions}

\author{Malihe Ghodrat}
\affiliation{School of Physics, Institute for Research in Fundamental Sciences (IPM), Tehran 19395-5531, Iran}

\author{Ali Naji}
\thanks{Email: \texttt{a.naji@ipm.ir}}
\affiliation{School of Physics, Institute for Research in Fundamental Sciences (IPM), Tehran 19395-5531, Iran}

\author{Haniyeh Komaie-Moghaddam}
\affiliation{School of Physics, Institute for Research in Fundamental Sciences (IPM), Tehran 19395-5531, Iran}

\author{Rudolf Podgornik}
\affiliation{Department of Theoretical Physics, J. Stefan Institute, SI-1000 Ljubljana, Slovenia}
\affiliation{Department of Physics, Faculty of Mathematics and Physics, University of Ljubljana, SI-1000 Ljubljana, Slovenia}


\begin{abstract}
We study the effective interaction mediated by strongly coupled Coulomb fluids between dielectric surfaces carrying quenched, random monopolar charges with equal mean and variance, both when the Coulomb fluid  consists only of mobile multivalent counterions and when it consists of an asymmetric ionic mixture containing multivalent and monovalent (salt) ions in equilibrium with an aqueous bulk reservoir. We analyze the consequences that follow from the interplay between surface charge disorder, dielectric and salt image effects, and the strong electrostatic coupling that results from multivalent counterions on the distribution of these ions and the effective interaction pressure they mediate between the surfaces. In a dielectrically homogeneous system, we show that the multivalent counterions are attracted towards the surfaces with a singular, disorder-induced potential that diverges logarithmically  on approach to the surfaces, creating a singular but integrable counterion density profile that exhibits an algebraic divergence at the surfaces with an exponent that depends on the surface charge (disorder) variance. This effect drives the system towards a state of lower thermal `disorder', one that can be described by a renormalized temperature, exhibiting thus a remarkable antifragility. In the presence of an interfacial dielectric discontinuity, the singular behavior of counterion density at the surfaces is removed but multivalent counterions are still accumulated much more strongly close to  randomly charged surfaces as compared with uniformly charged ones. The interaction pressure acting on the surfaces displays in general a highly non-monotonic behavior as a function of the inter-surface separation with a prominent regime of attraction at small to intermediate separations. This attraction is caused directly by the combined effects from charge disorder and strong coupling electrostatics of multivalent counterions, which dominate the surface-surface repulsion due to the (equal) mean charges on the two surfaces and the osmotic pressure of monovalent ions residing between them. These effects can be quite significant even with a small degree of surface charge disorder relative to the mean surface charge. The strong coupling, disorder-induced attraction is typically far more stronger than the van der Waals interaction between the surfaces, especially within a range of several nanometers for the inter-surface separation, where such effects are predicted to be most pronounced.
\end{abstract}


\maketitle

\section{Introduction}

The counter-intuitive electrostatic effects produced by mobile multivalent counterions in the vicinity of charged macromolecular surfaces (such as biopolymers like DNA, membranes, colloids, nano-particles and virus-like nano-capsids) have been studied extensively within the counterion-only models and in the case of non-disordered, homogeneous surface charge distributions  (see, e.g.,  recent reviews in Refs. \cite{holm,hoda_review, perspective, Naji_PhysicaA, Levin02, Shklovs02,French-RMP,book} and references therein). These effects include, most notably, the so-called like-charge attraction induced by multivalent counterions between charged surfaces of the same sign. The like-charge attraction phenomena stand at odds with the {\em weak coupling} (WC) paradigm based traditionally on the mean-field Poisson-Boltzmann (PB) theory \cite{VO,Israelachvili,Safranbook,Hunter,andelman-rev} and requires the framework of the {\em strong coupling} (SC) theories  \cite{book,holm,hoda_review, perspective, Naji_PhysicaA,Shklovs02,Levin02,Netz01,AndrePRL,AndreEPJE,asim,Naji_epje04,trizac} that incorporate strong ion-surface correlations, to the leading order, and ion-ion correlations, to subleading orders, ubiquitous in the situation where electrostatic interactions are strong; this is realized at low temperatures, solvents with low dielectric constant, and/or with highly charged macromolecular surfaces but, most prominently, with multivalent counterions. Usually, however, charged bio-soft systems with multivalent counterions include also a finite amount of monovalent salt ions that couple weakly to charged surfaces. A few experimental examples of such systems are furnished by the condensation of DNA by multivalent cations in the bulk \cite{Bloom2, Yoshikawa1, Yoshikawa2,Pelta, Plum,Raspaud,Pelta2} or in  viruses and virus-like nano-capsids \cite{Savithri1987,deFrutos2005,Siber}, and formation of large aggregates (or bundles) of highly charged biopolymers such as DNA \cite{rau-1,rau-2},
F-actin \cite{Angelini03} and microtubules \cite{Needleman}. In these situations, one deals with a difficult problem in which neither the WC  nor the SC  limiting laws can be applied without reservations. For such asymmetric Coulomb fluids, a generalized {\em dressed multivalent-ion approach}, that bridges the two limits in one single theoretical framework, has been introduced  \cite{SCdressed1,perspective} and tested against extensive explicit- and implicit-ion simulations  \cite{SCdressed1,SCdressed2,SCdressed3,perspective,leili1,leili2}.

In counterion-only systems and in the case of uniformly charged surfaces with surface charge density $-\sigma e_0$, the strength of the electrostatic interactions is described by the {\em electrostatic coupling parameter} $\Xi=q^2 \lB / \mu = 2\pi q^3\lB^2\sigma$  \cite{Netz01}. It is the ratio  of the rescaled Bjerrum length, $q^2\lB$ with $\lB=e_0^2/ (4 \pi \epsilon \epsilon_0k_{\mathrm{B}}T)$,  that measures   the strength of electrostatic interactions between  counterions of charge valency $q$, and the Gouy-Chapman length, $\mu=1/(2\pi\lB q \sigma)$, that measures  the strength of electrostatic interaction between a counterion and the surface charge in the units of thermal energy $k_B T$ (conventionally, we take $\sigma>0$ and $q>0$). In the WC regime, $\Xi\ll 1$,   typically realized with monovalent counterions, thermal energy is dominant  and, therefore, the counterions form a diffuse gas-like phase next to a charged surface that can be described by the mean-field PB theory. In the SC regime, $\Xi\gg 1$, typically realized with multivalent counterions,  electrostatic interactions are large enough to reduce the three-dimensional ionic cloud above the charged surface to a quasi-two-dimensional, strongly correlated liquid layer, or even a 2D crystal \cite{holm,book,hoda_review, perspective, Naji_PhysicaA, Levin02, Shklovs02,Netz01,AndrePRL,AndreEPJE,trizac}.

Subsequent studies also revealed that the WC-SC paradigm can be generalized to more complicated and realistic situations usually encountered in experimental context, with an asymmetric Coulomb fluid comprising  a mixture of mono- and multivalent  ions in the vicinity of charged macromolecular interfaces. Neither the WC nor the SC paradigm can be consistently applied to this situation where the system is electrostatically part strongly and part weakly coupled. In this case {\em dressed multivalent-ion theory}, based on piecewise application of the WC and SC formalism, has been shown to posses a robust regime of validity  \cite{SCdressed1,SCdressed2,SCdressed3,perspective,leili1,leili2}. Other cases demanding an extension and/or modification of the WC-SC framework have been recently reviewed  in Ref. \cite{perspective}.

A separate issue, of the nature of the charge distribution on macromolecular surfaces, has also received progressively more focused attention. In many cases, these surfaces are not only inhomogeneous in terms of the surficial charge configuration, but exhibit a fundamentally disordered distribution of charges \cite{manne1,manne2,Meyer,Meyer2,klein,klein1,klein2,science11,andelman-disorder,kantor-disorder0,Ben-Yaakov-dis,netz-disorder,netz-disorder2,Lukatsky1,Lukatsky2,Rabin,Miklavic,ali-rudi,rudiali,partial,disorder-PRL,jcp2010,pre2011,epje2012,jcp2012,book,jcp2014,speake,kim2,kim3,safran1,safran2,safran3,Olvera0,Hribar} dictated either by the method of sample preparation and/or by the inherent structural properties of the surface materials \cite{manne1,manne2,Meyer,Meyer2,klein,klein1,klein2,science11,kim2,kim3}. These disordered systems, with either thermalized, {\em annealed} surface charges \cite{netz-disorder,netz-disorder2,andelman-disorder,disorder-PRL,jcp2010,safran1,safran2,safran3,Olvera0,Miklavic} or structurally disordered, {\em quenched} surface charges \cite{andelman-disorder,kantor-disorder0,Ben-Yaakov-dis,netz-disorder,netz-disorder2,Lukatsky1,Lukatsky2,Rabin,Miklavic,ali-rudi,rudiali,partial,disorder-PRL,jcp2010,pre2011,epje2012,jcp2012,book,jcp2014,speake,kim2,kim3}  (or even partially annealed or partially quenched ones \cite{partial,Hribar}) have received  a more rigorous attention, as it is becoming clear that the nature of charge disorder can interfere fundamentally with the behavior of the system approximated with fixed, homogeneous surface charges.

In the most realistic model of the interaction between charged macromolecules, one would thus need to consider not only an asymmetric confined Coulomb fluid between macromolecular surfaces, with a mixture of mono- and multivalent salt ions, but also surface charges that exhibit a disordered component. While off hand this seems to be like a formidable task, detailed calculations have shown that the final results are rather intuitive and revealing \cite{ali-rudi,rudiali,disorder-PRL,jcp2010,pre2011,epje2012,jcp2012,jcp2014,partial}. First of all, it became clear that, for {\em quenched} charge disorder, which is of interest in this paper, and in the absence of dielectric asymmetry between the solution subphase and the macromolecular surface, there is no disorder-induced  effect in the WC regime \cite{ali-rudi,rudiali}. However, in the SC limit and in the same general regime of parameters, one finds a long-ranged,  disorder-induced attraction, which must be added to the repulsive force between equally charged surfaces \cite{ali-rudi}. It has also been shown that in the linear WC regime with Gaussian field-fluctuations around the mean-field solution, the quenched  disorder would lead to  repulsive or attractive interactions provided that the system has an inhomogeneous dielectric constant or monovalent salt distribution (repulsion arises in this case when the medium is more polarizable than the surfaces and vice versa) \cite{rudiali,disorder-PRL,jcp2010,pre2011,epje2012,jcp2012}.

Just as the coupling parameter $\Xi$ quantifies the electrostatic interactions between mobile counterions and fixed, homogeneous surface charges, the {\em disorder coupling parameter}, $\chi= 2\pi q^2 \lB^2 g$, quantifies the strength of the disorder-induced effects in the case of disordered surfaces with $g$ being the surface charge variance within the Gaussian {\em Ansatz} for its statistical distribution \cite{ali-rudi,rudiali,partial}. Obviously, the two coupling parameters, $\Xi$ and $\chi$, are very similar but the latter depends in a noticeably less drastic manner on the valency of the counterions. Just as the overall features of the behavior of a system with homogeneous charge distributions depend on the value of $\Xi$, the behavior of a disordered system depends in a very fundamental manner not only on the exact value of $\chi$, but also on the presence of  salt ions and dielectric inhomogeneities (``image charges") in the system \cite{disorder-PRL,jcp2010,pre2011,epje2012,jcp2012,jcp2014}.

Here, we  study the effective interaction mediated by strongly coupled Coulomb fluids between dielectric surfaces carrying quenched, random monopolar charges with equal mean and variance, both when the Coulomb fluid  consists only of mobile multivalent counterions and when it consists of an asymmetric ionic mixture. We analyze the consequences that follow from the interplay between surface charge disorder, dielectric and salt image effects, and the strong electrostatic coupling that result from multivalent counterions on the distribution of these ions and the effective interaction pressure they mediate between the surfaces. In a dielectrically homogeneous system,  the multivalent counterions are found to be attracted towards the surfaces with a singular, disorder-induced potential that diverges logarithmically on approach to the surfaces, creating a singular but integrable counterion density profile that exhibits an algebraic divergence at the surfaces with an exponent given by the disorder coupling parameter. This effect drives the system towards a state of lower thermal `disorder', one that can be described by a renormalized effective temperature, exhibiting thus a remarkable antifragility \cite{taleb} reported in a previous work by the present authors \cite{jcp2014}. We thus find  that, in this general context of strongly coupled, disordered systems, anti-fragile  behavior is ubiquitous and stems from the interplay of structural and thermal disorder, with the former  engendering a decrease in the {\em translational entropy} of multivalent counterions as a consequence of the fact that the disordered charge distribution generates a finite degree of (non-thermal) {\em configurational entropy}. In the presence of an interfacial dielectric discontinuity, the singular behavior of the counterion density at the surfaces is removed but multivalent counterions are still accumulated much more strongly close to  randomly charged surfaces as compared with uniformly charged ones. The interaction pressure between the surfaces displays in general a highly non-monotonic behavior as a function of the inter-surface separation with a prominent regime of attraction at small to intermediate separations. This attraction is caused directly by the combined effects from charge disorder and SC electrostatics of multivalent counterions, which dominate the surface-surface repulsion due to the (equal) mean charges on the two surfaces and the osmotic pressure of monovalent salt ions residing between them. These effects are quite significant even with a small degree of surface charge disorder (variance) relative to the mean surface charge.

The organization of the paper is as follows: We introduce our model in Section \ref{sec:model} and present our theoretical formalism in Section \ref{sec:formal}. The distribution of multivalent counterions in the disordered counterion-only case is studied in Section \ref{subsec:ci-only} and the effects of salt ions and dielectric inhomogeneities are considered in Section \ref{subsec:Density_salt}. The behavior of the effective inter-surface pressure is studied in Sections \ref{subsec:P}-\ref{subsec:dis}, followed by the conclusion and discussion in Section \ref{sec:discussion}.

\section{The Model}
\label{sec:model}

\begin{figure}[t!]
\begin{center}
\includegraphics[width=8.5cm]{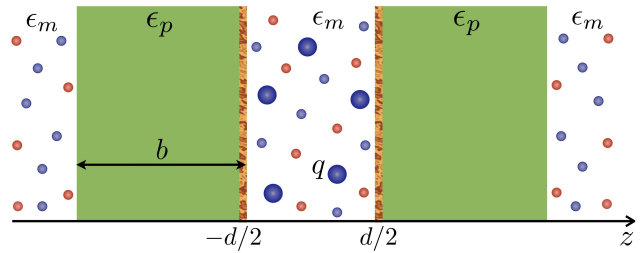}
\caption{(Color online) Schematic view of two infinite, plane-parallel dielectric slabs of thickness $b$ and dielectric constant $\epsilon_p$ in a bathing ionic solution of dielectric constant $\epsilon_m$, containing a monovalent and multivalent salt mixture (with multivalent counterions shown by large blue spheres and monovalent salt anions and cations shown by small orange and blue spheres). The multivalent counterions are confined in the slit between the randomly charged inner surfaces of the slabs. }
\label{fig:schematic}
\end{center}
\end{figure}

We consider two plane-parallel dielectric slabs of infinite surface area $S$, finite thickness $b$ and dielectric constant $\epsilon_p$, placed perpendicular to the $z$ axis with the inner bounding surfaces separated by a distance  $d$ (see Fig. \ref{fig:schematic}). The outer bounding surfaces of the slabs are assumed to be neutral, while the inner ones carry a {\em quenched}, spatially uncorrelated, random charge distribution $\rho(\Av r)$, characterized by a Gaussian probability weight
\begin{equation}
   {\mathcal P}[\rho] = C\, \rme^{- \frac{1}{2}  \int \!{\mathrm{d}} {\mathbf r}\,  g^{-1}({\mathbf r})\,
   								[\rho({\mathbf r}) - \rho_0({\mathbf r}) ]^2}.
\label{eq:pdf}								
\end{equation}
Here, 
$C$ is a normalization factor and $\rho_0({\mathbf r})$ and $g({\mathbf r})$ are the mean and the variance of the disordered charge distribution, respectively. For the specific model considered here, we assume that the charge distribution on the inner slab surfaces are statistically identical and given by
\begin{eqnarray}
	\rho_0({\mathbf r}) &=& -\sigma e_0 \, \big[\delta(z+d/2)+\delta(z-d/2)\big], \\
	g({\mathbf r}) &=& g e_0^2 \, \big[\delta(z+d/2)+\delta(z-d/2)\big],
\end{eqnarray}
where, with no loss of generality, we take $g\geq 0$ and  $\sigma> 0$  (in general, the surfaces may also have no net charge but only a finite disorder variance, which we shall consider in detail elsewhere \cite{preprint3}). It should be noted that, we develop the main formalism for an arbitrary shape of the charged boundaries, which we then apply to the specific example of the two planar slabs  described above.

The  slabs are assumed to be immersed in a solution of dielectric constant $\epsilon_m$  containing an asymmetric Coulomb fluid  composed of  a monovalent $1:1$ salt of bulk concentration $n_0$ and a multivalent $q:1$ salt of bulk concentration $c_0$ with $q > 0$ being the charge valency of multivalent counterions. The  dielectric slabs are assumed to be impermeable to mobile ions. The multivalent counterions are assumed to be confined within the slit $-d/2\leq z\leq d/2$. This particular constraint enables us to reproduce the standard  counterion-only results \cite{ali-rudi} as a limiting  case (practically, multivalent counterions  can be prevented from entering the outer regions, $|z|>d/2+b$, by enclosing these regions in semi-permeable membranes). This assumption has no impact on our results in a large part of the parameter space, e.g., for slab thicknesses larger than the Debye screening length, and especially for semi-infinite slabs that will be of main interest in this work.

\section{The Formalism}
\label{sec:formal}

\subsection{Dressed multivalent-ion theory}

The problem of an asymmetric Coulomb fluid containing monovalent and multivalent ions next to charged macromolecular surfaces is a complicated one, mainly because different components of the Coulomb fluid couple differently to the surface charges: Multivalent counterions tend to couple strongly, and therefore generate non-mean-field effects, whereas monovalent anions and cations couple weakly, and are thus expected to follow the standard mean-field paradigms 
as they would in the absence of multivalent ions \cite{VO,Israelachvili,Safranbook,Hunter,andelman-rev}. Strong coupling effects due to multivalent counterions within the counterion-only models can be described well using the recent SC theories that have been studied thoroughly over the last several years  (see, e.g.,  Refs. \cite{holm,hoda_review, perspective, Naji_PhysicaA, Levin02, Shklovs02,French-RMP,book} and references therein). In the case of an asymmetric Coulomb fluid, although it is not {\em a priori} clear if a single approximation can be introduced in order to describe the hybrid nature of the electrostatic couplings in the presence of charged boundaries, it turns out that a simple generalization of the counterion-only SC theory \cite{hoda_review, perspective, Naji_PhysicaA,Netz01,AndrePRL,AndreEPJE,asim,Naji_epje04}, dubbed {\em dressed multivalent-ion theory},  provides a very good approximation in a large part of the parameter space as discussed in detail in Refs.   \cite{SCdressed1,SCdressed2,SCdressed3,perspective,leili1,leili2}. This approach is obtained by first tracing the partition function over the degrees of freedom associated with monovalent ions within a linearization approximation and then virial expanding it with respect to the fugacity of multivalent counterions. Both these steps can be justified systematically \cite{SCdressed1} only in the case of highly asymmetric mixtures with a sufficiently large counterion valency $q$ and provided that the bulk concentration of multivalent counterions is relatively small (as is often the case in experimental systems containing asymmetric ionic mixtures  \cite{Bloom2, Yoshikawa1, Yoshikawa2,Pelta, Plum,Raspaud,Pelta2,Savithri1987,deFrutos2005,Siber,rau-1,rau-2,Angelini03,Needleman}); because otherwise the different ionic species in the solution can not be treated on such different levels. 
The dressed multivalent-ion theory described and reviewed recently \cite{perspective} bridges the gap between the weak coupling DH theory and the counterion-only SC theory and reproduces them as two special limiting cases at large and small screening parameters, respectively.

\subsection{General form of the interaction free energy}

In the dressed multivalent-ion theory, the grand-canonical partition function of an asymmetric Coulomb fluid with a fixed realization of external charges  $\rho({\mathbf r})$ can be written, in a functional-integral form \cite{Edwards,Podgornik89,Podgornik89b,Netz-orland}, as follows \cite{SCdressed3}
\begin{equation}
  {\mathcal Z}[\rho] = \rme^{-\frac{1}{2}\ln \,\det\, G} \int \!{\mathcal D}\phi \,\, \rm\rme^{-\beta S[\phi, \rho]}.
  \label{eq:Z}
\end{equation}
Here, $\beta = 1/(k_{\mathrm{B}}T)$, $\phi({\mathbf r})$ is the fluctuating electrostatic potential and $S[\phi, \rho]$ is the effective ``field-action", i.e.,
\begin{eqnarray}
\label{fieldaction}
  &&S[\phi, \rho] =  \frac{1}{2} \int \rmd {\mathbf r}\,  \rmd {\mathbf{r}'} \phi(\mathbf{r}) G^{-1}({\mathbf r}, {\mathbf r}') \phi(\mathbf{r}') +\\
  &&\qquad\quad + \,\rmi \int \rmd {\mathbf r}\,\rho({\mathbf r})\phi({\mathbf r}) -  k_{\mathrm{B}}T\lambda_c \int \rmd {\mathbf r}\,\Omega_c({\mathbf r})  \rm\rme^{-\rmi\beta q e_0  \phi},\nonumber
\end{eqnarray}
where $\lambda_c=c_0$ is the fugacity (or  bulk concentration) of multivalent counterions, $\Omega_c({\mathbf r})$ is an indicator function that determines the region of space that is accessible to multivalent counterions (e.g., in the present model, we have  $\Omega_c({\mathbf r})= \theta(z+d/2)-\theta(z-d/2)$), and $G^{-1}({\mathbf r}, {\mathbf r}')$ is the operator inverse of the Green's function $G({\mathbf r}, {\mathbf r}')$ that, in the dressed multivalent-ion theory, satisfies the DH equation
\begin{equation}
-\epsilon_0 \nabla\cdot \epsilon({\mathbf r}) \nabla G({\mathbf r}, {\mathbf r}')  + \epsilon_0  \epsilon({\mathbf r})\kappa^2({\mathbf r})G({\mathbf r}, {\mathbf r}')= \delta({\mathbf r} -{\mathbf r}').
\label{eq:G_DH}
\end{equation}
Here, $\kappa({\mathbf r})$ is the Debye (or salt) screening parameter which is non-zero only outside the region occupied by the dielectric slabs, i.e., $\kappa^2 = 4\pi \ell_{\mathrm{B}} n_b$,  
where $n_b = 2n_0+qc_0$ is the bulk concentration due to all monovalent ions. Therefore, the role of monovalent ions is incorporated into the Green's function on the DH level.

As noted before, in the presence of multivalent counterions with sufficiently small bulk concentration, the partition function can be virial-expanded in terms of the counterion fugacity, $\lambda_c$, as
\begin{equation}
\label{eq:firstordr_Gpar}
\mathcal{Z}[\rho]=\mathcal{Z}_0[\rho]+\lambda_c \mathcal{Z}_1[\rho]+\mathcal{O} (\lambda_c^2),
\end{equation}
with the first two terms corresponding to the dressed multivalent-ion theory on the leading order. The zeroth-order term corresponds to the electrostatic interaction of fixed charged objects  in the absence of multivalent counterions,
\begin{equation}
\label{eq:Gpar_zero}
\mathcal{Z}_0[\rho]= \rme^{-\frac{1}{2}\ln \,\det\, G-\frac{\beta}{2} \int \!{\mathrm{d}}\Av r \, {\mathrm{d}}\Av r'\, \rho(\Av r)G(\Av r,\Av r')\rho(\Av r')},
\end{equation}
and the first-order term gives the single-particle contribution due to multivalent counterions
\begin{equation}
\label{eq:Gpar_one}
\nonumber \mathcal{Z}_1[\rho]= \mathcal{Z}_0[\rho]\!\int\! {\mathrm{d}}\Av r \,\Omega(\Av r) \,\rme^{-\beta u({\mathbf r}; [\rho])}.
\end{equation}
Here, $u({\mathbf r}; [\rho])$ is the single-particle interaction energy that has the form
\begin{equation}
  u({\mathbf r}; [\rho]) = qe_0 \int {\mathrm{d}}{\mathbf r}' \,G({\mathbf r}, {\mathbf r}')\rho({\mathbf r}') + \frac{q^2e_0^2}{2} G_{\mathrm{im}}({\mathbf r}, {\mathbf r}),
  \label{eq:u_rho}
\end{equation}
where $G_{\mathrm{im}}({\mathbf r}, {\mathbf r})$ is the generalized Born energy contribution, which is generated purely by the dielectric and/or salt polarization effects (or the  ``image charges"), i.e., $G_{\mathrm{im}}({\mathbf r}, {\mathbf r})= G({\mathbf r}, {\mathbf r}) - G_0({\mathbf r}, {\mathbf r})$, with  $G_0({\mathbf r}, {\mathbf r})$ representing the formation (self-)energy of individual counterions in a homogeneous background; this can be calculated from the screened free-space  Green's function defined through $-\epsilon_0 \epsilon_m(\nabla^2 - \kappa^2) G_0({\mathbf r}, {\mathbf r}')  = \delta({\mathbf r} -{\mathbf r}')$.

In the case of counterion-only systems, the two leading terms in Eq. (\ref{eq:firstordr_Gpar}) were shown to generate a finite contribution to the free energy, constituting the SC theory \cite{hoda_review, perspective, Naji_PhysicaA,Netz01,AndrePRL,AndreEPJE,asim,Naji_epje04}. The higher-order terms contain subleading contributions from multi-particle interactions that become important only in the crossover regime between the WC and SC regimes. In the more general context of dressed multivalent-ion theory, the leading-order virial terms can generate both the SC and WC behaviors of the system in the limits of small and large screening parameters, respectively \cite{SCdressed1}. The predictions of this theory were analyzed for uniformly charged surfaces (and also for strictly neutral surfaces) and were compared with extensive numerical simulations elsewhere \cite{SCdressed1,SCdressed2,SCdressed3,perspective,leili1,leili2}.

In the present study, the charge distribution of the dielectric surfaces, $\rho(\Av r)$, has a quenched, Gaussian disordered component with a finite variance  around the mean charge density $\rho_0(\Av r)$.  One therefore needs to average the thermodynamic quantities over different realizations of the charge disorder as well. Hence, for instance, the free energy follows from
\begin{equation}
\beta {\mathcal  F}= -\langle\!\langle\ln\, \mathcal{Z}[\rho]\rangle\!\rangle.
\end{equation}
The disorder average is given by  $\langle\!\langle \cdots \rangle\!\rangle=\int \mathcal{D}\rho \, (\cdots) {\mathcal P}[\rho]$ and the Gaussian weight by Eq. (\ref{eq:pdf}).  The averaged quantity $\langle\!\langle\ln \mathcal{Z}[\rho]\rangle\!\rangle$  can be calculated in general by employing the Edwards-Anderson's replica {\em Ansatz} \cite{ali-rudi}.
Since we are interested  only in the leading virial terms, we can also directly average the leading-order free energy terms over the quenched charge disorder weight, yielding
\begin{equation}
\label{eq:firstorder_Gpot}
\beta {\mathcal  F}= -\langle\!\langle\ln\, \mathcal{Z}_0\rangle\!\rangle+\lambda_c \langle\!\langle\frac{\mathcal{Z}_1}{\mathcal{Z}_0}\rangle\!\rangle+{\mathcal O}(\lambda_c^2).
\end{equation}
The averages in the above expression can be carried out straightforwardly and we find the (grand-canonical) dressed multivalent-ion free energy of the system as
\begin{eqnarray}
\label{eq:exact_Gpot}
&&\beta {\mathcal  F}=\frac{\beta}{2} \int {\mathrm{d}}\Av r\, {\mathrm{d}}\Av r'\,\rho_0(\Av r)G(\Av r,\Av r')\rho_0(\Av r')\\
\nonumber &&\qquad +\frac{\beta}{2}\mathrm{Tr}\big[g(\Av r)G(\Av r,\Av r')\big]- \lambda_c\int {\mathrm{d}}\Av R\, \Omega(\Av R)\, \rme^{-\beta u(\Av R)},\nonumber
\end{eqnarray}
where the first and the second terms are the contributions in the absence of multivalent counterions, representing  the interaction free energies due to the mean surface charge density, $\rho_0(\Av r)$, and the disorder variance, $g(\Av r)$, respectively. The second term arises also in the analysis of the fluctuation-induced forces between disordered surfaces in vacuum or in a weakly coupled Coulomb fluid \cite{rudiali,disorder-PRL,jcp2010,pre2011,epje2012,jcp2012}; it gives a non-vanishing contribution only in inhomogeneous systems with a finite dielectric discontinuity at the bounding surfaces and/or a spatially inhomogeneous distribution of salt ions. The third term represents the contribution from multivalent counterions on the leading (single-particle) level, in which $u({\mathbf r})$ is
the {\em effective} single-particle interaction energy \cite{jcp2014}
\begin{eqnarray}
  &&u({\mathbf r}) = qe_0 \int {\mathrm{d}}{\mathbf r}'\, G({\mathbf r}, {\mathbf r}')\rho_0({\mathbf r}')+\frac{q^2e_0^2}{2} G_{\mathrm{im}}({\mathbf r}, {\mathbf r})\nonumber \\
   &&\qquad - \beta  \frac{q^2e_0^2}{2} \int {\mathrm{d}}{\mathbf r}' g(\Av r) [G(\Av r,\Av r')]^2.
   \label{eq:u2}
\end{eqnarray}
We note that the first term in Eq. (\ref{eq:u2}) originates from the interaction of multivalent counterions with the mean surface charge density, the second term  from the self-interactions of individual counterions (with their own image charges) and the third term is due to the presence of surface charge disorder. This term is proportional to the disorder variance and shows an explicit temperature dependence and a quadratic dependence on the Green's function and the multivalent ion charge valency, $q$; it arises from the sample-to-sample fluctuations (or variance) of the single-particle interaction energy (\ref{eq:u_rho})  \cite{jcp2014}.

In the SC limit or within the multivalent dressed-ion theory \cite{SCdressed1,Netz01,AndreEPJE}, the density profile of multivalent counterions can be derived solely in terms of the effective single-particle interaction energy as
\begin{equation}
	\label{eq:sc_density_av}
         c({\mathbf r})  =    \lambda_c \Omega({\mathbf r}) \, \rme^{-\beta u({\mathbf r})}.
\end{equation}
This concludes the recapitulation of the multivalent dressed-ion theory that forms the basic framework of our analysis of electrostatic coupling and quenched  charge disorder in what follows.

\subsection{Specific case of planar dielectric slabs}
\label{subsec:slabs}

The free energy expression (\ref{eq:exact_Gpot}) is valid regardless of the shape of the boundaries. In the rest of this paper, however, we shall delimit ourselves to the specific example of two planar dielectric slabs (Section \ref{sec:model}), in which case the Green's function  $G(\Av r,\Av r') = G(\Avg\rho, \Avg\rho';z, z')$, with $\Avg\rho = (x, y)$ and $\Avg\rho' = (x', y')$ being the transverse (in-plane) coordinates, is only a function of  $|\Avg\rho-\Avg\rho'|$. Thus, its  Fourier-Bessel transform $\hat G( Q;z,z')$ can be defined by
\begin{equation}
G(\Av r,\Av r') = \int_0^\infty \frac{Q {\mathrm{d}}Q}{2\pi }\,\hat G(Q;z,z')\,J_0(Q \vert \Avg\rho -\Avg\rho'\vert).
\end{equation}
Using standard methods, we find
\begin{eqnarray}
\nonumber &&\hat G(Q;z,z')=\frac{1}{2\epsilon_0\epsilon_m \gamma }[e^{-\gamma |z-z'|}+
\frac{2 e^{-2\gamma d}\Upsilon(Q b)}{1-e^{-2\gamma d}\Upsilon^2(Q b)}\\
\nonumber &&\qquad\times(e^{\gamma d}\cosh{\gamma(z+z')}+\Upsilon(Q b) \cosh{\gamma(z-z')})],\\
\label{eq:G_finite_b}
\end{eqnarray}
where $\gamma^2=\kappa^2+Q^2$,
\begin{equation}
\Upsilon(Q b)=\frac{\Delta_s(1-\rme^{-2Q b})}{1-\Delta_s^2 \,\rme^{-2 Q b}},
\end{equation}
and
\begin{equation}
\Delta_s=\frac{\epsilon_m \gamma-\epsilon_p Q }{\epsilon_m \gamma +\epsilon_p Q}.
\end{equation}
For semi-infinite slabs ($b\rightarrow \infty$), one can recover the well-known expression
\begin{eqnarray}
\label{eq:G_infinite_b}
 &&\hat G(Q;z,z')= \frac{1}{2\epsilon_0\epsilon_m \gamma }\big[\rme^{-\gamma |z-z'|}+
\frac{2\Delta_s \,\rme^{-2\gamma d}}{1-\Delta_s^2\,\rme^{-2\gamma d}}\\
\nonumber &&\qquad\qquad\times\big(\rme^{\gamma d}\cosh{\gamma(z+z')}+\Delta_s \cosh{\gamma(z-z')}\big)\big].
\end{eqnarray}
Note that the information about the image charge effects, which result from the inhomogeneous distribution of the dielectric constant or the bathing salt solution in the system, enter here through the parameter $\Delta_s$. In the absence of salt ions ($\kappa=0$), $\Delta_s$ reduces to the bare dielectric discontinuity parameter
\begin{equation}
\label{eq:dielec_discontinuity}
\Delta= \frac{\epsilon_m-\epsilon_p}{\epsilon_m +\epsilon_p}.
\end{equation}
In the cases treated below, relevant for  aqueous solvents in the presence of bounding slabs of low dielectric constant, we delimit ourselves to $\Delta\geq 0$, which gives {\em repulsive}  image interactions.

The effective single-particle interaction energy in the two-slab system can be written using Eq. (\ref{eq:u2})  as
\begin{eqnarray}
\label{eq:ResU_neutral}
 u(z)&=&-q \sigma e_0^2\, \big[\hat G(0;z,-d/2)+ \hat G(0;z, d/2)\big]\\
\nonumber        &&+\frac{q^2 e_0^2}{4\pi} \int_0^\infty Q \rmd Q\, \hat G(Q; z,z)\\
\nonumber && -\frac{\beta q^2  g e_0^4}{4\pi} \!\int_0^\infty\! \!Q \rmd Q \big[\hat G^2(Q;z, -d/2)+\hat  G^2(Q;z,d/2)\big].
\end{eqnarray}
The density profile of counterions in the slit region, $-d/2\leq z\leq d/2$, then follows as  (see Eq. (\ref{eq:sc_density_av}))
\begin{equation}
	\label{eq:sc_density_av_2}
         c(z)  =    \lambda_c \, \rme^{-\beta u(z)},
\end{equation}
and the grand-canonical free energy  (per $k_{\mathrm{B}}T$ and per surface area, $S$) can  be written as
\begin{eqnarray}
\label{eq:Free energy}
&&\frac{\beta {\mathcal F}}{S}=\frac{4\pi \ell_{\mathrm{B}}\sigma^2}{ \kappa}\coth(\kappa d/2)\nonumber\\
&&\qquad\quad+\,g \ell_{\mathrm{B}}f(\kappa, d, \Delta)-\lambda_c \int_{-d/2}^{d/2} \rmd z\, \rme^{-\beta u(z)}.
\end{eqnarray}
In the above expression, the different contributing terms  appear in the same order as in Eq. (\ref{eq:exact_Gpot}) and we have
\begin{equation}
\label{eq:f}
f(\kappa, d, \Delta)\equiv  \int_{0}^{\infty} Q \rmd Q \,\frac{\Delta_s(1+\Delta_s)^2}{\gamma(\rme^{2d\gamma}-\Delta_s^2)},
\end{equation}
where we have omitted irrelevant additive terms that are independent of the surface separation, $d$. In fact, the {\em interaction} free energy follows from Eq. (\ref{eq:Free energy}) by subtracting all such terms or, in other words, the reference free energy of the system for $d\rightarrow \infty$. 

\subsection{Dimensionless representation}

It is convenient to make use of a dimensionless  set of quantities by rescaling the spatial coordinates with the Gouy-Chapman length $\mu = 1/(2\pi q \ell_{\mathrm{B}}  \sigma)$ as $\tilde{\mathbf r}={\mathbf r}/\mu$. Other parameters are rescaled accordingly, e.g., the inter-surface separation, the salt screening parameter and an analogously defined length scale, $\chi_c^2 = 8\pi q^2 \ell_{\mathrm{B}} c_0$, which is referred to as  the rescaled bulk concentration of multivalent counterions, i.e.,
\begin{equation}
\tilde d = d/\mu, \quad \tilde\kappa=\kappa\mu, \quad  \tilde \chi_c = \chi_c\mu.
\end{equation}
The rescaling of the Bjerrum length leads to the dimensionless electrostatic coupling parameter $\Xi=q^2 \lB / \mu= 2\pi q^3 \lB^2 \sigma$, associated with the mean surface charge density \cite{Netz01}, while the surface charge variance appears in the dimensionless  disorder coupling (or  strength) parameter,  $\chi= 2\pi q^2 \lB^2 g$  \cite{ali-rudi}. The rescaling of the density profile and the interaction pressure will be discussed later.

\section{Results}

\subsection{Distribution of counterions: The counterion-only model}
\label{subsec:ci-only}

Let us first consider the distribution of multivalent counterions in the special case of the counterion-only model, where the salt screening and image charge effects are set to zero by assuming a homogeneous system with $\kappa=0$ and $\Delta=0$, or equivalently, $\epsilon_p=\epsilon_m$ (this case was considered in a previous work  \cite{ali-rudi}, which however focused only on the effective interaction between the surfaces and did not investigate the distribution of counterions).  The two impermeable, randomly charged  surfaces are placed at a separation distance, $d$, confining in the slit between them a fixed number of multivalent counterions, $N$, which is fixed by the mean charge on the two surfaces through the global electroneutrality condition,  $Nq=2S\sigma$. Hence, the fugacity of counterions is given by \cite{ali-rudi,Netz01}
 \begin{equation}
\lambda_c =\frac{N}{S \int_{-d/2}^{d/2} {\mathrm{d}}z\, \rme^{-\beta u(z)}}.
\end{equation}
 The Green's function of the counterion-only model reduces to the bare Coulomb interaction, $G_0({\mathbf r}, {\mathbf r}') = 1/(4\pi \epsilon_0\epsilon_m |{\mathbf r} - {\mathbf r}'|)$, and  the effective single-particle interaction (up to an irrelevant additive constant and in rescaled units) follows straightforwardly from Eqs. (\ref{eq:u2}) or (\ref{eq:ResU_neutral}) as
\begin{equation}
  \beta u(\tilde z) =  \tilde d + \frac{\chi}{2}\ln \bigg(\frac{\tilde d^2}{4}-\tilde z^2\bigg).
   \label{eq:u2_ci_only}
\end{equation}
Note that while, because of the symmetric plane-parallel geometry of the model, there is no net attraction acting on individual counterions from the (oppositely signed) mean surface charges, the counterions experience an attractive potential due to the quenched surface charge disorder, which is given by the second term in Eq. (\ref{eq:u2_ci_only}).

The rescaled density profile of counterions in the slit is obtained by using Eq. (\ref{eq:sc_density_av_2}) as
 \begin{equation}
\tilde c(\tilde z)\equiv\frac{c(\tilde z)}{2\pi \ell_{\mathrm{B}} \sigma^2}= \bigg(\frac{2}{\tilde d}\bigg)C_0^{-1}(\chi)\bigg(\frac{1}{4}-\frac{\tilde z^2}{\tilde d^2}\bigg)^{-\chi/2},
\label{eq:dens_0}
\end{equation}
where we have defined
\begin{equation}
  C_0(\chi) = 2^{-1+\chi}\sqrt{\pi}\,\frac{\Gamma\big(1-\frac{\chi}{2}\big)}{\Gamma\big(\frac{3}{2}-\frac{\chi}{2}\big)}.
\end{equation}

\begin{figure}[t!]\begin{center}
	\begin{minipage}[h]{0.32\textwidth}\begin{center}
		\includegraphics[width=\textwidth]{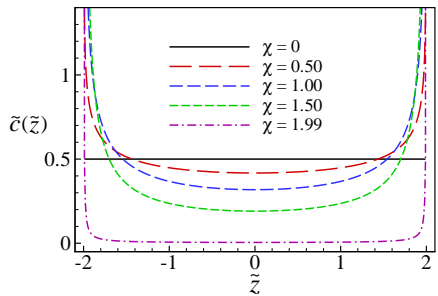}
	\end{center}\end{minipage}
\caption{(Color online) Rescaled density profiles of  multivalent counterions within the counterion-only model as a function of the rescaled normal position, $\tilde z=z/\mu$, between two randomly charged surfaces located at inter-surface separation $\tilde d=d/\mu = 4$. Different curves correspond to different values of the disorder coupling parameter as shown on the graph.}
\label{fig:inf_dist_0}
\end{center}
\end{figure}

In the no-disordered case ($\chi=0$), the above expression reduces to the  standard SC density profile  $\tilde c(\tilde z)=2/\tilde d$  \cite{Netz01,AndreEPJE,AndrePRL,hoda_review,perspective,Naji_PhysicaA,Shklovs02}. As shown in Fig. \ref{fig:inf_dist_0},  the  density profile of multivalent counterions is strongly modified by the surface charge disorder  and exhibits algebraic singularities close to the randomly charged boundaries at $z=\pm d/2$ for any finite value of the disorder coupling parameter, in clear violation of the contact-value theorem established for uniformly charged surfaces \cite{contact_value,contact_value2,contact_value1,contact_value3}. This kind of behavior was discussed in our recent work on the distribution of multivalent counterions next to a single, randomly charged interface \cite{jcp2014} and we shall not delve further into the details, making only a few remarks in what follows.

The singular behavior of the counterion density profile comes directly from the logarithmic disorder term in the single-particle energy  in Eq.  (\ref{eq:u2_ci_only}), which drives the counterions towards the surfaces. The density of counterions away from the surfaces decreases as the disorder coupling parameter increases  (Fig. \ref{fig:inf_dist_0}) and, eventually, it tends to zero $\tilde c(\tilde z)\rightarrow 0$ for any $|\tilde z|< \tilde d/2$ when $\chi\rightarrow 2^-$ since $C_0(\chi)\rightarrow \infty$. Counterions are thus densely accumulated in the immediate vicinity of the surfaces. This behavior must be distinguished from the surface adsorption or counterion condensation phenomena \cite{perspective} as, in the present context, one can show systematically that the mean surface charge is not renormalized by the surface accumulation of counterions  (see Appendix C in Ref. \cite{partial}).

The accumulation of counterions in the vicinity of the charged boundaries furthermore drives the system towards a state of lower thermal `disorder', since, as one can show, the {\em translational entropy} of the multivalent counterions decreases as a consequence of the disordered charge distribution generating a finite (non-thermal) {\em configurational entropy}; this latter type of entropy stems from the different realizations of the quenched disorder. This behavior was associated in Ref. \cite{jcp2014} with the {\em anti-fragility} \cite{taleb} of the system, in which introducing an external (quenched) disorder source diminishes the intrinsic thermal disorder and drives the system  towards a more `ordered' state, which
 is also a thermodynamically more stable one as compared with the corresponding non-disordered case (see Ref. \cite{ali-rudi} for the canonical free energy expression of the disordered counterion-only model).

The rescaled interaction pressure acting on each of the surfaces, $\tilde P = \beta P/(2\pi \ell_{\mathrm{B}} \sigma^2)$, can be obtained as \cite{ali-rudi}
 \begin{equation}
\tilde P= -1+\frac{2(1-\chi)}{\tilde d}.
\label{eq:P_SC}
\end{equation}
which, by setting $\chi=0$, reduces to the standard SC pressure between two non-disordered like-charged surfaces \cite{Netz01}
\begin{equation}
\tilde P_0= -1+\frac{2}{\tilde d}.
\label{eq:P_SC_0}
\end{equation}
The attractive electrostatic and repulsive entropic terms in the pressure give rise to an equilibrium (rescaled) bound-state separation of $\tilde d_*=2$ between two identical, uniformly charged surfaces \cite{Netz01,AndreEPJE,AndrePRL,hoda_review, Naji_PhysicaA}. As is clear from Eq. (\ref{eq:P_SC}), the surface charge disorder gives an additive attractive contribution to the total interaction pressure that renormalizes the entropic  term and leads to a more closely packed bound state with the equilibrium inter-surface separation $\tilde d_*= 2(1-\chi)$. This separation tends to zero for $\chi\rightarrow 1^-$, beyond which the two surfaces collapse into a primary minimum, because  the entropic contribution (second term in Eq. (\ref{eq:P_SC})) changes sign at $\chi=1$ and, hence, the pressure becomes attractive at all separations (and diverges for  $\tilde d\rightarrow 0$) in the regime $\chi> 1$.

Another notable point is that the renormalization of the repulsive entropic term by a factor $1-\chi$ can be interpreted also as a renormalization of the effective temperature of the system to a lower value.
However, this interpretation should be considered with caution because first, this particular form of the renormalization of the repulsive pressure is found only in the SC limit of the {\em two-surface counterion-only} model and, secondly, one can show, by inspecting the thermodynamic quantities of the system,  that the renormalizing term, $-2\chi/\tilde d$, in Eq. (\ref{eq:P_SC}) indeed comes from both electrostatic energy and entropic contributions.

\begin{figure*}[t!]\begin{center}
	\begin{minipage}[h]{0.32\textwidth}\begin{center}
		\includegraphics[width=\textwidth]{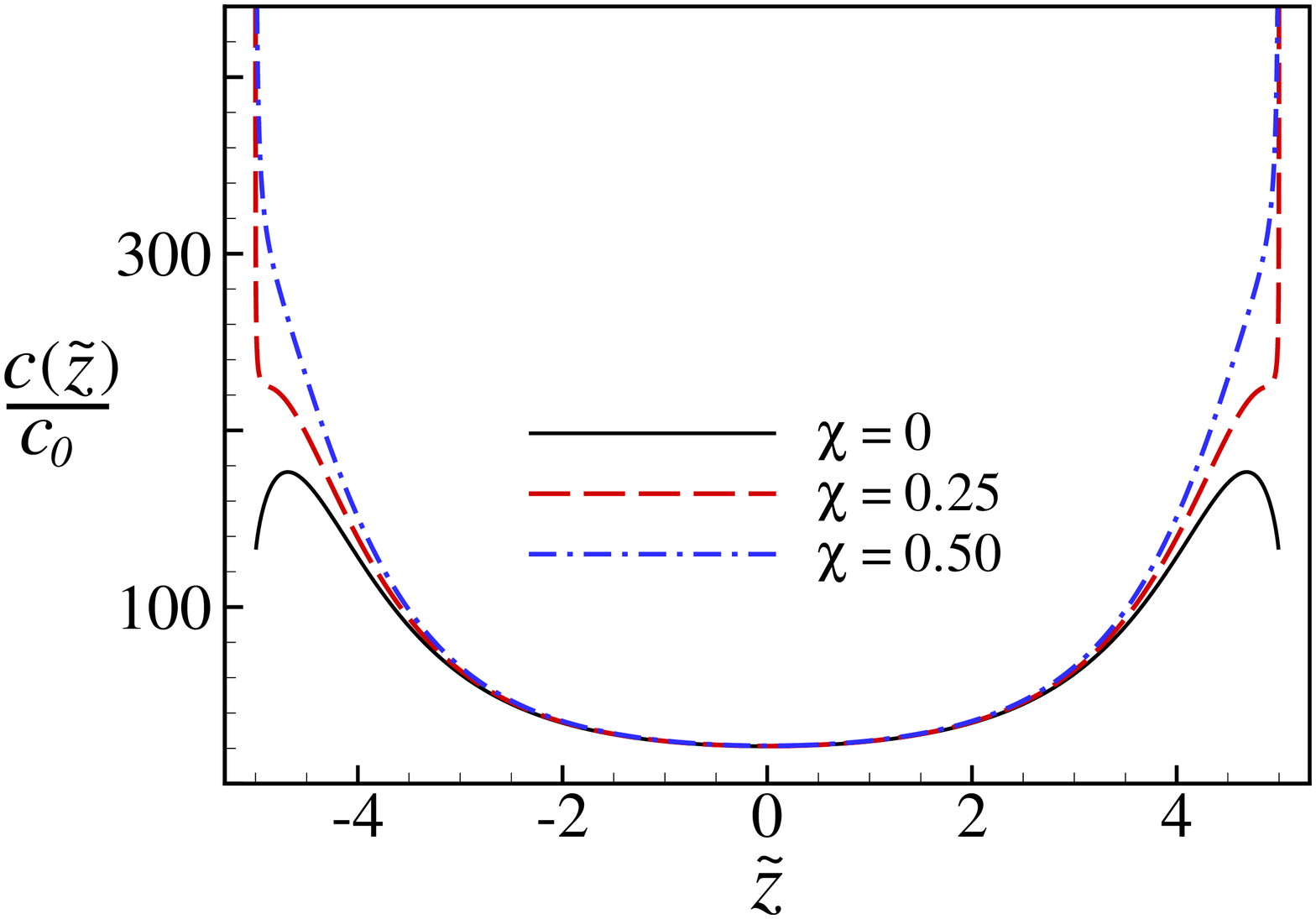} (a)
	\end{center}\end{minipage} \hskip0.2cm	
	\begin{minipage}[h]{0.32\textwidth}\begin{center}
		\includegraphics[width=\textwidth]{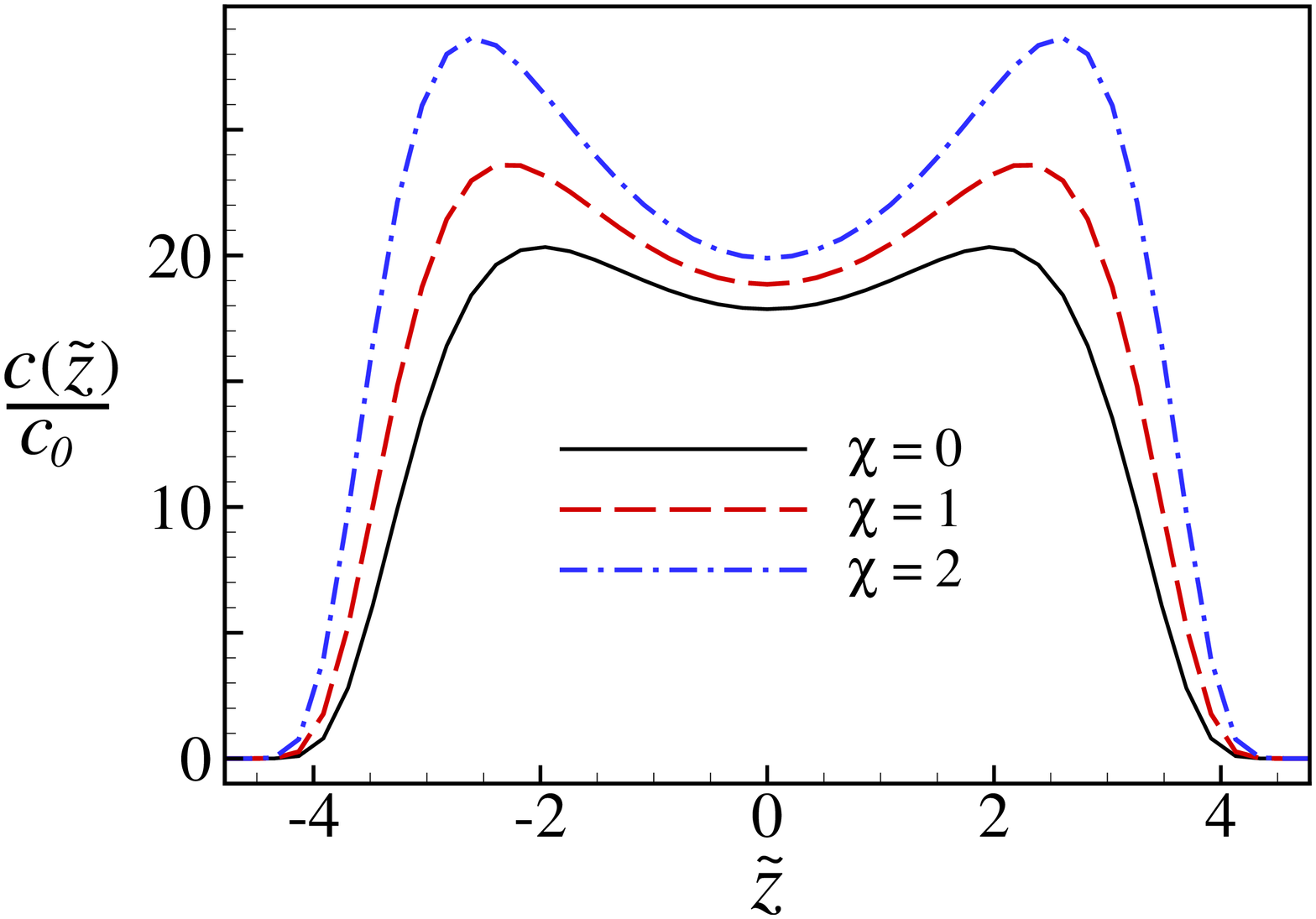} (b)
	\end{center}\end{minipage} \hskip0.2cm	
    \begin{minipage}[h]{0.32\textwidth}\begin{center}
		\includegraphics[width=\textwidth]{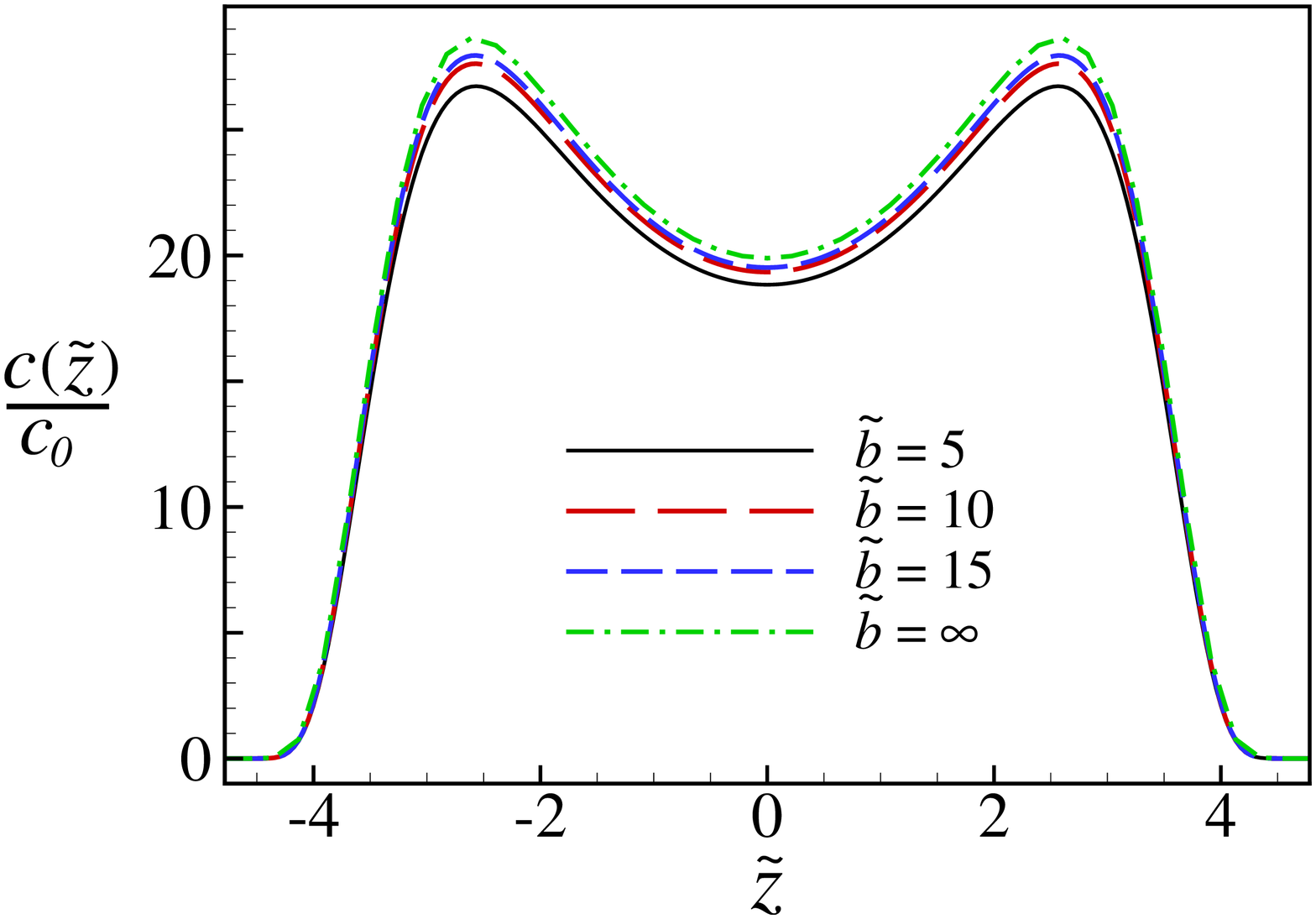}  (c)
	\end{center}\end{minipage} 	
\caption{(Color online) Rescaled density profiles of dressed multivalent counterions as a function of the rescaled normal position, $\tilde z$, between the randomly charged inner surfaces of two semi-infinite slabs placed at rescaled separation $\tilde d=10$ in (a) dielectrically homogeneous ($\Delta=0$) and (b) dielectrically inhomogeneous systems ($\Delta=0.95$) at fixed parameter values $\tilde \kappa=0.3$ and $\Xi =50$ and various disorder coupling parameters as shown on the graphs. (c) is the same as (b) but with  fixed $\chi=2$ and varying  slab thickness  $\tilde b=5$, 10, 15 and $\infty$.}
\label{fig:inf_dist}
\end{center}\end{figure*}

\subsection{Distribution of counterions: Salt screening and image charge effects}
\label{subsec:Density_salt}

In the next step, we assume that, in addition to the multivalent counterions that are introduced through an asymmetric $q:1$ salt of bulk concentration  $c_0$, the ionic mixture also contains monovalent salt of bulk concentration $n_0$. As noted before, the Debye screening parameter is $\kappa = (4\pi \ell_{\mathrm{B}} n_b)^{1/2}$ with   $n_b = 2n_0+qc_0$. For the time being, we also assume that the system is  dielectrically homogeneous, i.e., $\epsilon_p=\epsilon_m$ and that the  slabs are  semi-infinite  ($ b =\infty$) and impermeable to all ions. The density profile of multivalent counterions can be calculated from Eqs.  (\ref{eq:ResU_neutral}) and (\ref{eq:sc_density_av_2})  and by making use of the appropriate expressions from Eqs. (\ref{eq:G_finite_b})-(\ref{eq:G_infinite_b}).

In the absence of surface charge disorder, the density profile of multivalent counterions shows a non-monotonic behavior with a peak at a small distance from each of the two bounding surfaces (see the black solid curve in  Fig. \ref{fig:inf_dist}a; note also that here the density profiles are rescaled with the bulk value $c_0$). This behavior is due to  the interplay between two distinct factors; namely, the salt screening effect and the ``salt image" effect. The former  dominates at  intermediate to large distances from the  surfaces that are  comparable to or larger than the Debye screening length $\tilde \kappa^{-1}$, while the latter dominates at small to intermediate distances from the surfaces, causing a partial  depletion of multivalent counterions from the proximity of the  surface boundaries; it  is generated because of the inhomogeneous distribution of salt ions that are not allowed to permeate into the wall regions (i.e., $|\tilde z|>\tilde d/2$), leading in turn  to a discontinuous change in the polarization of the medium at the interfacial boundaries, with the slit region having  a larger polarizability response than the slab region. This gives rise to repulsive  ``salt image" effects that show some qualitative similarities to the ``dielectric image" effects.

When the bounding surfaces carry a finite degree of quenched charge randomness ($\chi>0$), the multivalent counterions  exhibit a strong attraction towards them and, again, generate a singular density profile with a diverging contact value (dashed curves  in Fig. \ref{fig:inf_dist}a). In fact, the  counterion density profile exhibits  the same algebraic singularity on approach to the surfaces as in  the counterion-only case, Eq. (\ref{eq:dens_0}) \cite{jcp2014}. Thus, at small distances from the surfaces, the disorder-induced effects overcome both salt image and salt screening effects.

These features change qualitatively when the system is dielectrically inhomogeneous and exhibits a dielectric discontinuity at the interfacial boundaries at $z=\pm d/2$. In Fig.  \ref{fig:inf_dist}b, we show the counterion density profiles for $\Delta=0.95$ (corresponding to the dielectric discontinuity at the water/hydrocarbon boundary with $\epsilon_m=80$ and $\epsilon_p=2$) and a few different values of the disorder coupling parameter.  Strong dielectric image repulsions lead to a zone of complete depletion  near the surfaces with vanishing counterion density, followed by  enhanced peaks  at an intermediate distance from each surface. These features are qualitatively similar in the presence or absence of surface charge disorder, although the  disorder generates larger densities, especially at the peak regions, by attracting a larger number of multivalent counterions from the bulk reservoir into the slit. The interfacial depletion zone is generated because (unlike the salt image effects) the dielectric image effects can be described in terms of point-like image charges (especially for large $\Delta\simeq 1$) \cite{SCdressed3}; these images repel the counterions with a singular image  potential (second term in Eqs. (\ref{eq:u2}) and (\ref{eq:ResU_neutral}))) that behaves approximately as the first-image interaction potential $\beta u_{\mathrm{im}} \simeq \Xi\Delta/4(\tilde d/2\pm \tilde z)$ at small distances from the boundaries, thus overcoming the logarithmic, disorder-induced  attraction experienced by the  counterions.

\begin{figure*}[t]\begin{center}
	\begin{minipage}[h]{0.32\textwidth}\begin{center}
		 \includegraphics[width=\textwidth]{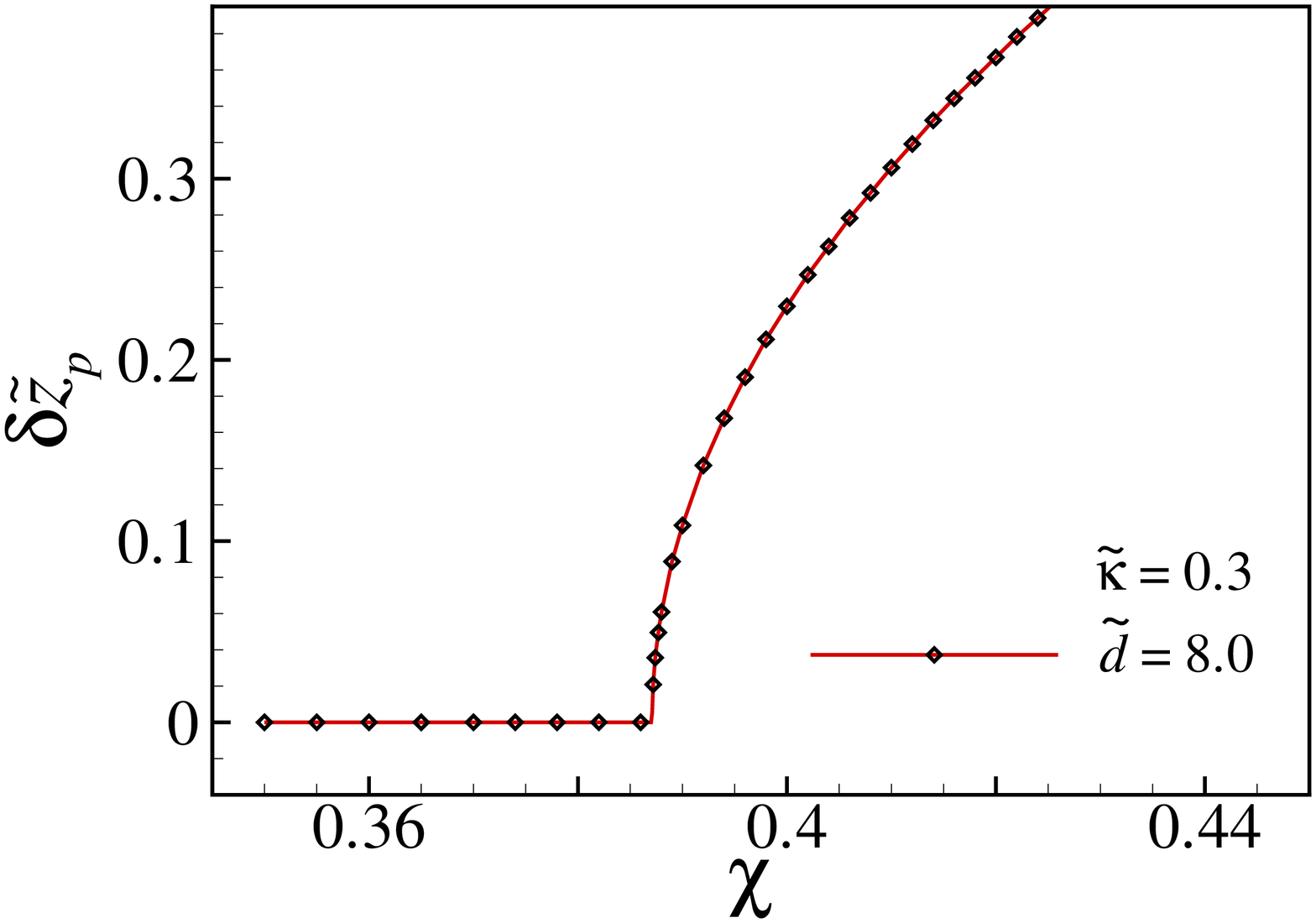} (a)
	\end{center}\end{minipage} \hskip0.2cm
	\begin{minipage}[h]{0.32\textwidth}\begin{center}
		\includegraphics[width=\textwidth]{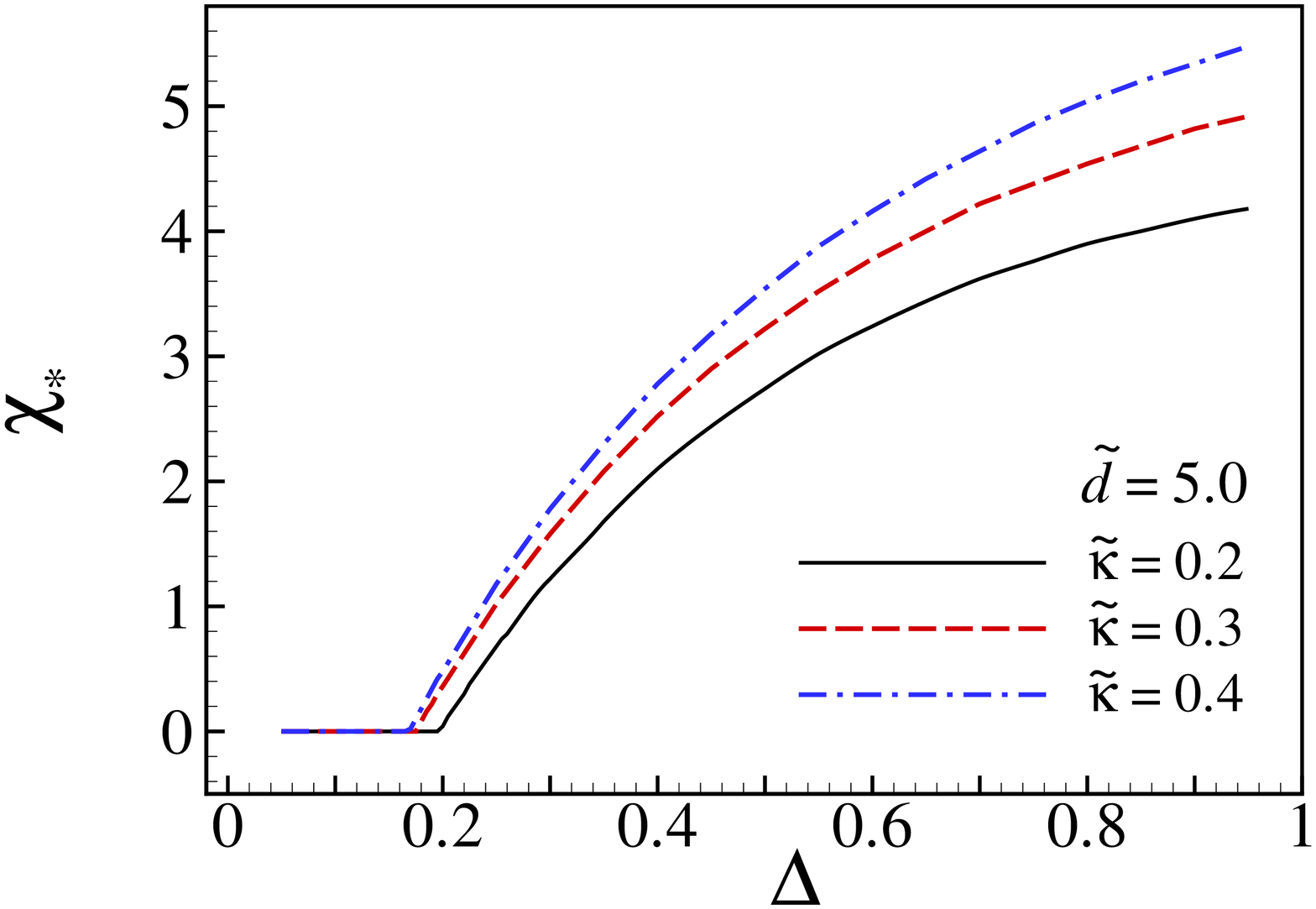} (b)
	\end{center}\end{minipage} \hskip0.2cm	
	\begin{minipage}[h]{0.32\textwidth}\begin{center}
		\includegraphics[width=\textwidth]{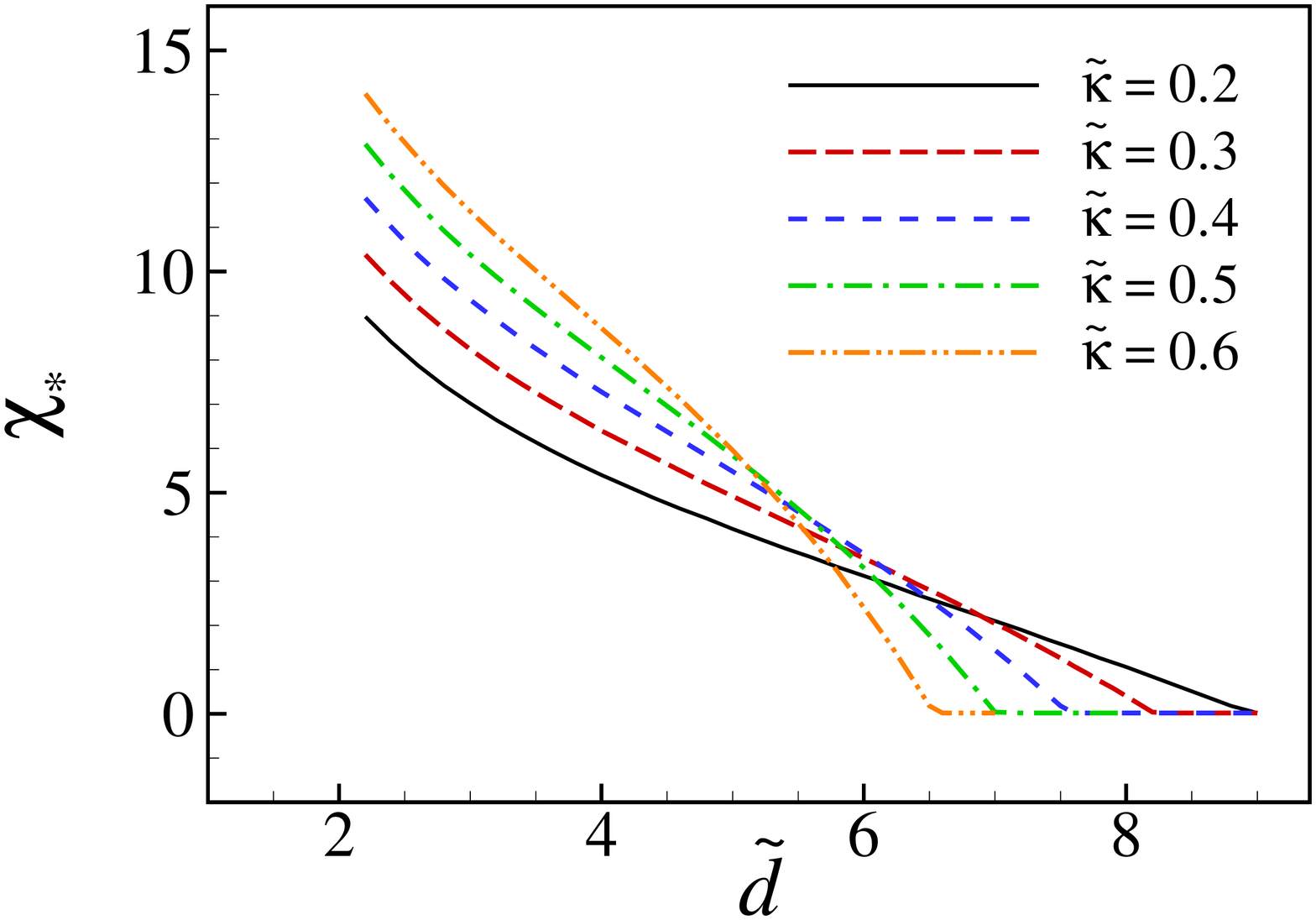} (c)
	\end{center}\end{minipage}
\caption{(Color online) (a) Rescaled distance between the peaks (humps) in the density profile of multivalent counterions as a function of the disorder coupling parameter, $\chi$, for $\tilde \kappa=0.3$,  $\tilde d=8$ and $\Delta = 0.95$.  (b) The threshold value $\chi_*$, where the counterion density profile changes from a unimodal to a bimodal shape, as a function of the dielectric discontinuity parameter, $\Delta$,  at fixed $\tilde d = 5$ and different values of $\tilde \kappa$ as shown on the graph. (c)  The threshold value $\chi_*$ as a function of the rescaled inter-surface distance, $\tilde d$,  at fixed $\Delta = 0.95$  and different values of $\tilde \kappa$. In (b) and (c), the region below (above) the curves  correspond to the parameter values for which the density profile is uni- (bi-) modal. In all these cases, the slabs are semi-infinite.
}
\label{fig:phase_diags}
\end{center}\end{figure*}

We should also note that the finiteness of the slab thickness has typically only a small effect on the counterion distribution, especially when it is comparable to or larger than the screening length, $ \kappa  b\gtrsim1$, which is in fact often the case in realistic systems. For instance, in Fig. \ref{fig:inf_dist}c, we show the results for the same parameters as in Fig. \ref{fig:inf_dist}b but with $\chi=2$ and different slab thicknesses in the range $\tilde b\geq 5$ (which covers the range $b>1$~nm in actual units, see the Discussions). The density of counterions in the slit is slightly increased but saturates quickly when the slab thickness is increased to infinity. The smaller counterion density found in the case of thinner slabs is due to the fact that the overall surface attraction experienced by counterions becomes smaller for smaller slab thicknesses
and, at the same time,   the salt ions in the outer  region  behind the slabs (see Fig. \ref{fig:schematic}) also contribute more strongly to the screening effects.

An interesting effect seen in the above results is that the competition between disorder-induced attraction and image repulsion leads to a highly pronounced bimodal profile with two humps that correspond to two distinct counterion-populated regions in the slit. (Such bimodal  profiles have also been found for the counterion density between heterogeneous but regularly patterned, planar charged surfaces \cite{Nikoofard} and also for the monomer density of polyelectrolyte chains between uniformly charged planar surfaces  \cite{rudi_PE}.) These humps are expected to appear  at sufficiently large disorder variances or relatively large inter-surface separations as compared with the screening length.
This kind of morphological change in the distribution of multivalent counterions  can be quantified by defining the distance between the peaks, $\delta z_p$,  as an analog of the `order parameter' in the phase transition context. As seen in Fig. \ref{fig:phase_diags}a (for $\tilde \kappa=0.3$, $\tilde d = 8$ and $\Delta=0.95$), this quantity  shows a sharp, continuous change at a threshold value of $\chi_*\simeq 0.385$ from a single-hump profile to a double-hump one. In Figs. \ref{fig:phase_diags}b and \ref{fig:phase_diags}c, we show the results obtained for the threshold value $\chi_*$ as a function of the dielectric discontinuity parameter, $\Delta$ (at fixed $\tilde d = 5$), and as a function of the rescaled inter-surface distance, $\tilde d$ (at fixed $\Delta=0.95$), respectively. The region below (above) the curves in these figures corresponds to the parameter values for which the density profile is uni- (bi-) modal. As seen in Fig. \ref{fig:phase_diags}b, at larger values of the dielectric discontinuity or at  larger salt screening parameters, a larger disorder coupling parameter (disorder variance) is required in order to counteract the counterion-image repulsions from the boundaries and create a bimodal structure. Same is true for a system with a smaller inter-surface separation, see Fig. \ref{fig:phase_diags}c.

The salt screening has a reverse effect at small or large inter-surface separations: Whereas at small separations it increases the values of $\chi_*$, at large separations it decreases them. Another point to be noted here is that, for the parameter values used in Figs. \ref{fig:phase_diags}b and \ref{fig:phase_diags}c, there is a plateau-like region with $\chi_*=0$, where  the density profile is bimodal for any value of $\chi$.

\subsection{Interaction pressure}
\label{subsec:P}

In the most general case, where the system is immersed in an asymmetric electrolyte bath, the interaction pressure (equivalent to the {\em osmotic} or {\em disjoining} pressure) acting on the slabs follows from the difference in the slit pressure and the bulk (electrolyte) pressure, i.e., $P=P_s-P_b$. The slit pressure is obtained by differentiating the free energy expression (\ref{eq:Free energy}) with respect to the inter-surface separation as $P_s = -{\partial {\mathcal F}}/({S\partial d})$, where all other parameters are kept fixed, and the bulk pressure is given by  $P_b =(n_b+c_0) k_{\mathrm{B}}T$ with $n_b = 2n_0+qc_0$ being the bulk concentration of the monovalent ions as defined before.
It should be noted that the slit pressure obtained from the differentiation of the dressed multivalent-ion free energy with respect to $d$  does not contain the contribution from the osmotic pressure of monovalent ions in the slit, $P_{mon}$, which can be calculated in terms of the mid-plane density of monovalent ions \cite{SCdressed3}.

\begin{figure*}[t!]\begin{center}
	\begin{minipage}[h]{0.32\textwidth}\begin{center}
		 \includegraphics[width=\textwidth]{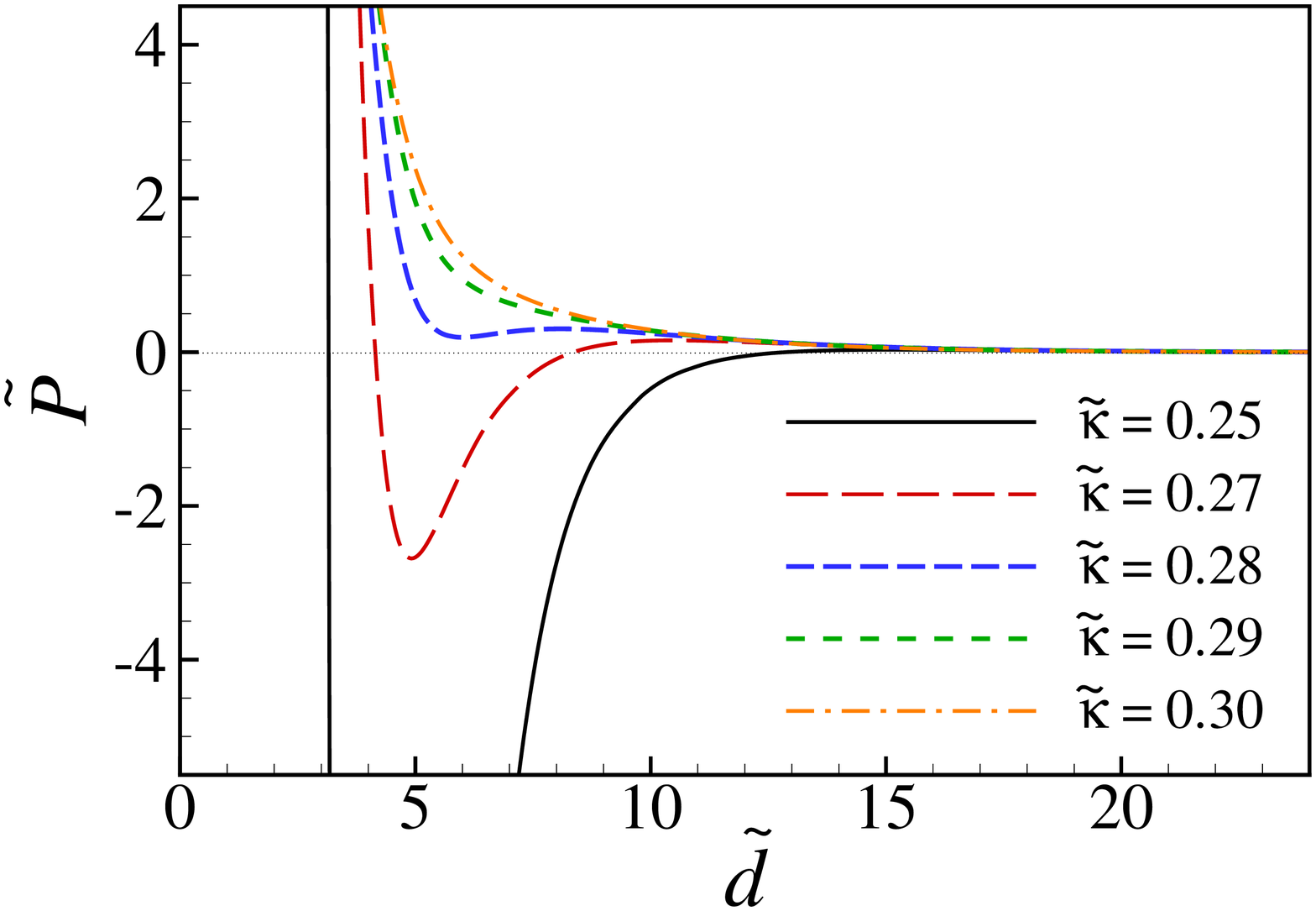} (a)
	\end{center}\end{minipage}
	\begin{minipage}[h]{0.32\textwidth}\begin{center}
		 \includegraphics[width=\textwidth]{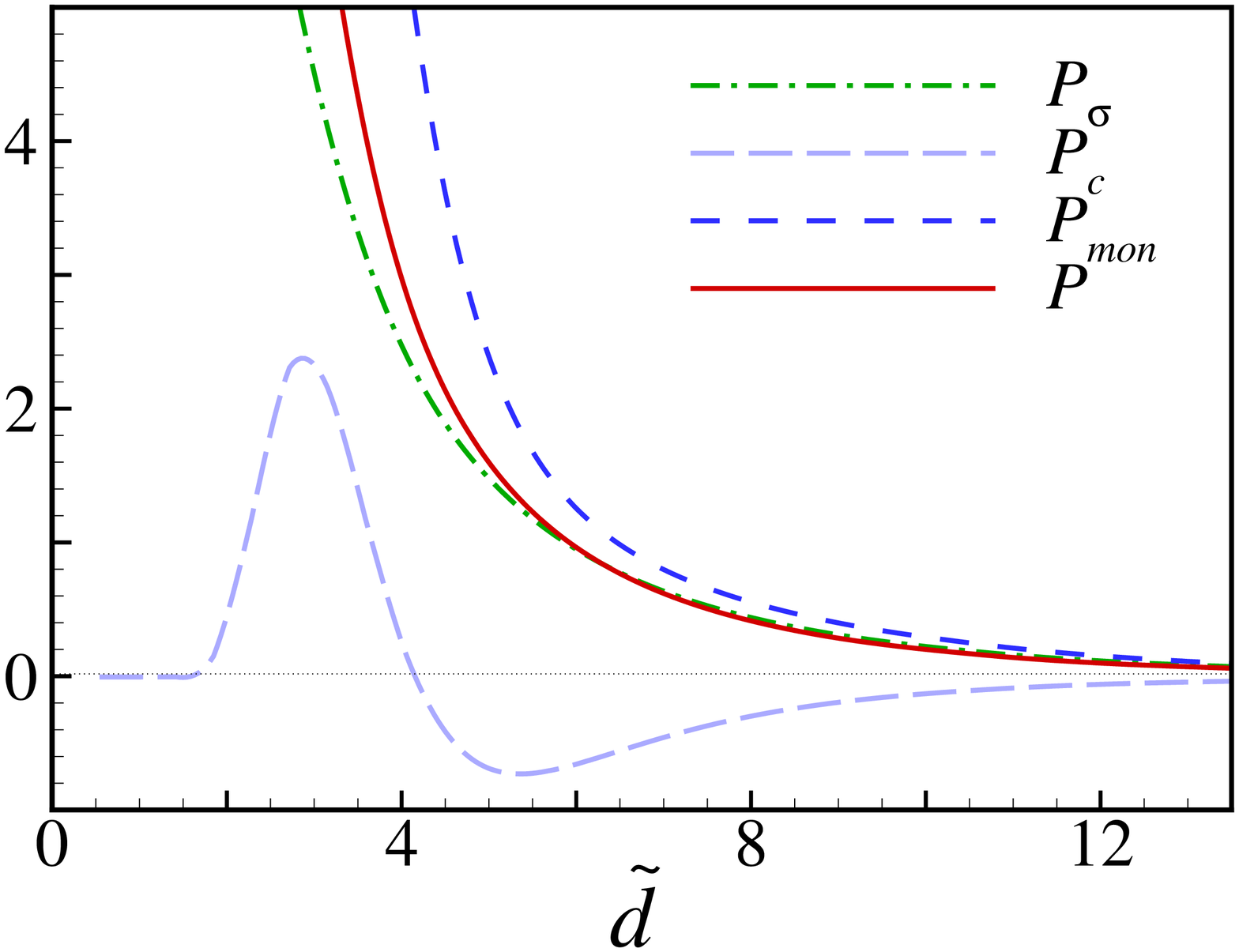} (b)
	\end{center}\end{minipage}
    \begin{minipage}[h]{0.32\textwidth}\begin{center}
		 \includegraphics[width=\textwidth]{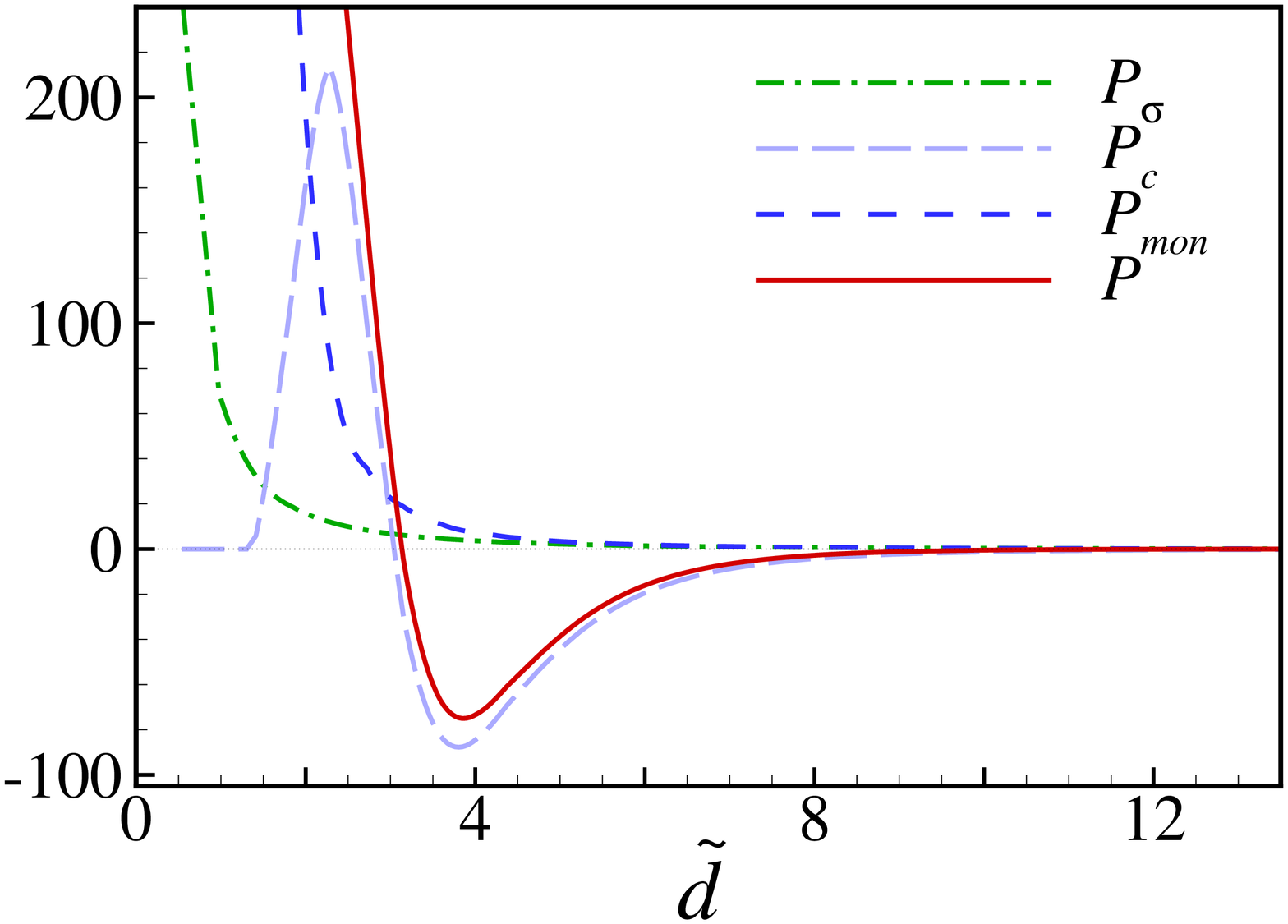} (c)
	\end{center}\end{minipage}
\caption{(Color online) (a) Rescaled interaction pressure as a function of the rescaled distance between the uniformly charged inner surfaces of two identical semi-infinite dielectric slabs  with  $\Xi=50$, $\Delta=0.95$ and $\tilde \chi_c=0.15$ and for a few different values of the rescaled screening parameter, $\tilde \kappa$, as indicated on the graph. Panels (b) and (c) show the different components that contribute to the interaction pressure  for  $\tilde \kappa=0.3$ and $\tilde \kappa=0.25$, respectively (see the text for definitions).}
\label{fig:pressure_no disorder}
\end{center}\end{figure*}

We decompose the rescaled interaction  pressure, $\tilde P = \beta P/(2\pi \ell_{\mathrm{B}} \sigma^2)$, into its different components as
\begin{equation}
\tilde P=\tilde P_\sigma+\tilde P_{dis}+\tilde P_{c}+\tilde P_{mon},
\label{eq:P_total}
\end{equation}
where
\begin{eqnarray}
\label{eq:P_sigma}
&&\tilde P_{\sigma}= \mathrm{csch}^2\bigg(\frac{\tilde \kappa \tilde d}{2}\bigg),\\
\label{eq:P_dis}
&&\tilde P_{dis}=-\chi\frac{\partial \tilde f(\tilde \kappa, \tilde d, \Delta)}{\partial \tilde d},\\
\label{eq:P_c}
&&\tilde P_{c}=\frac{\tilde \chi_c^2}{4}\bigg[\frac{\partial}{\partial \tilde d}\int_{-\tilde d/2}^{\tilde d/2} \rmd \tilde z \, \rme^{-\tilde u(\tilde z)} -1\bigg],\\
\label{eq:P_mon}
&&\tilde P_{mon}=\frac{q^2\tilde \kappa^2}{2}\big(\tilde n(0)-1\big).
\end{eqnarray}

The first term in Eq. (\ref{eq:P_total}), $\tilde P_{\sigma}$, follows from the first term in Eq. (\ref{eq:Free energy}) and represents the  repulsive pressure between the mean charges on the inner surfaces of the slabs. The second term, $\tilde P_{dis}$,  gives the contribution from the surface charge disorder (second term in Eq. (\ref{eq:Free energy})) with $\tilde f(\tilde \kappa, \tilde d, \Delta)$ being defined using Eq. (\ref{eq:f}) as
\begin{equation}
\label{eq:res_f}
\tilde f(\tilde\kappa, \tilde d, \Delta) \equiv  \int_{0}^{\infty} \tilde Q \rmd \tilde Q \,\frac{\Delta_s(1+\Delta_s)^2}{\tilde \gamma(\rme^{2\tilde d\tilde \gamma}-\Delta_s^2)},
\end{equation}
where $\tilde Q=Q\mu$ and $\tilde \gamma = \gamma \mu$. This  contribution can be attractive or repulsive and will be non-vanishing only in inhomogeneous systems with a finite dielectric discontinuity and/or an inhomogeneous distribution of salt ions (i.e., when $\Delta_s\neq 0$). These two contributions to the interaction pressure will be present regardless of the multivalent counterions; they have been analyzed in detail for semi-infinite slabs in Refs.  \cite{rudiali,disorder-PRL,jcp2010,pre2011,epje2012,jcp2012}.

The contributions $\tilde P_{c}$ and $\tilde P_{mon}$, on the other hand,  represent the osmotic pressure components from multivalent counterions and monovalent salt ions, respectively. $\tilde P_{c}$ is given in terms of the effective single-particle interaction energy, $\tilde u(\tilde z)$, and follows from the  third term in Eq. (\ref{eq:Free energy}) with  $\tilde u(\tilde z)$  obtained straightforwardly by rescaling the parameters  in Eq. (\ref{eq:ResU_neutral}).

Finally,  $\tilde P_{mon}$ is calculated from the contact-value theorem in terms of the total mid-plane density of monovalent ions, which can be estimated here through the relation $n(z)=\lambda_+ \exp[-\beta u(z)]\big|_{q=1}+\lambda_- \exp[-\beta u(z)]\big|_{q=-1}$ as discussed in detail in Ref. \cite{SCdressed3}, where $\lambda_+=n_0$ and $\lambda_- = n_0+qc_0$ are the bulk concentrations of monovalent cations and anions, respectively; the rescaled mid-plane density in Eq. (\ref{eq:P_mon}) is then defined as $\tilde n(0)=n(\tilde z=0)/n_b$.

\begin{figure*}[t!]\begin{center}
	\begin{minipage}[h]{0.32\textwidth}\begin{center}
		 \includegraphics[width=\textwidth]{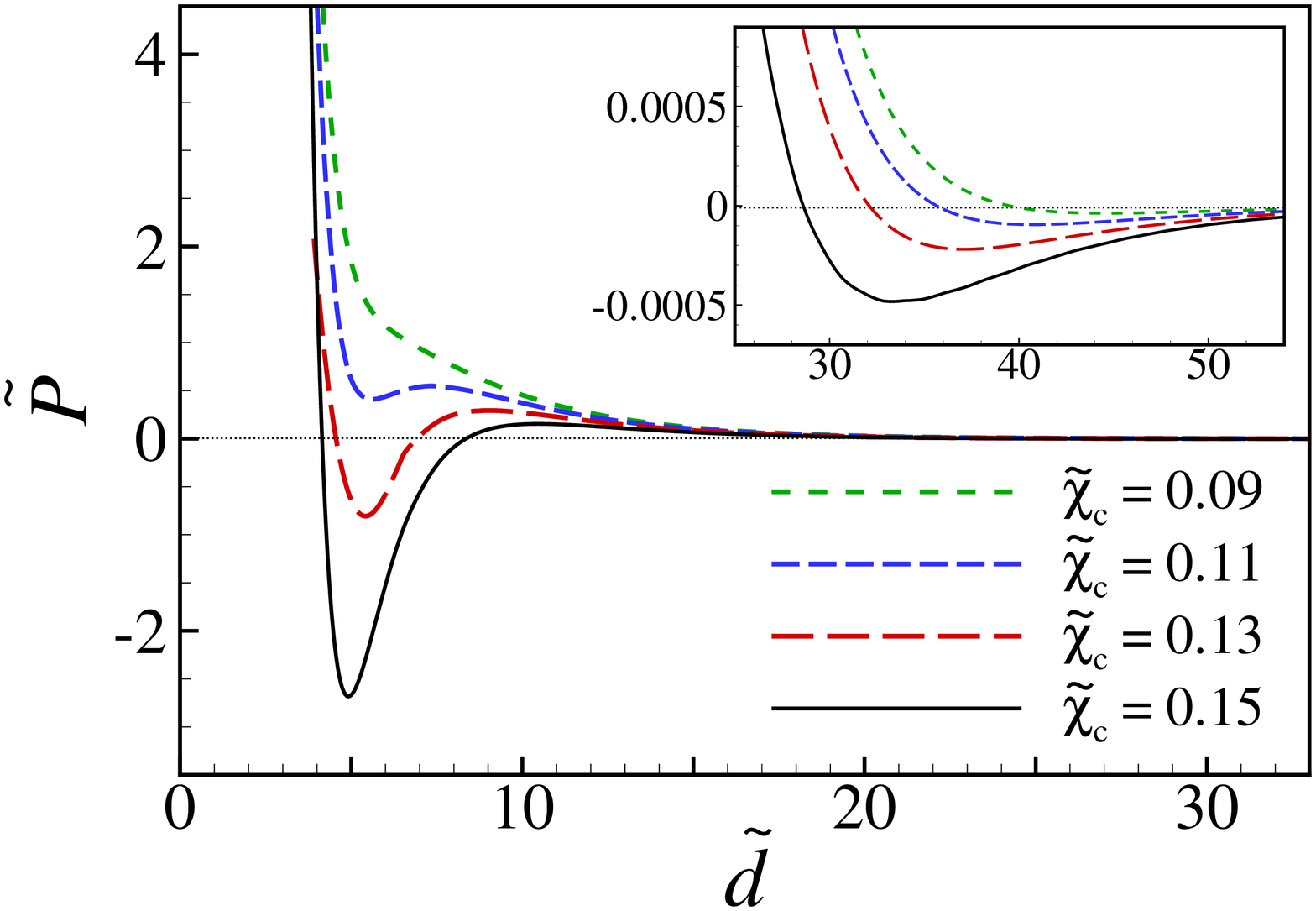} (a)
	\end{center}\end{minipage}
	\begin{minipage}[h]{0.32\textwidth}\begin{center}
		 \includegraphics[width=\textwidth]{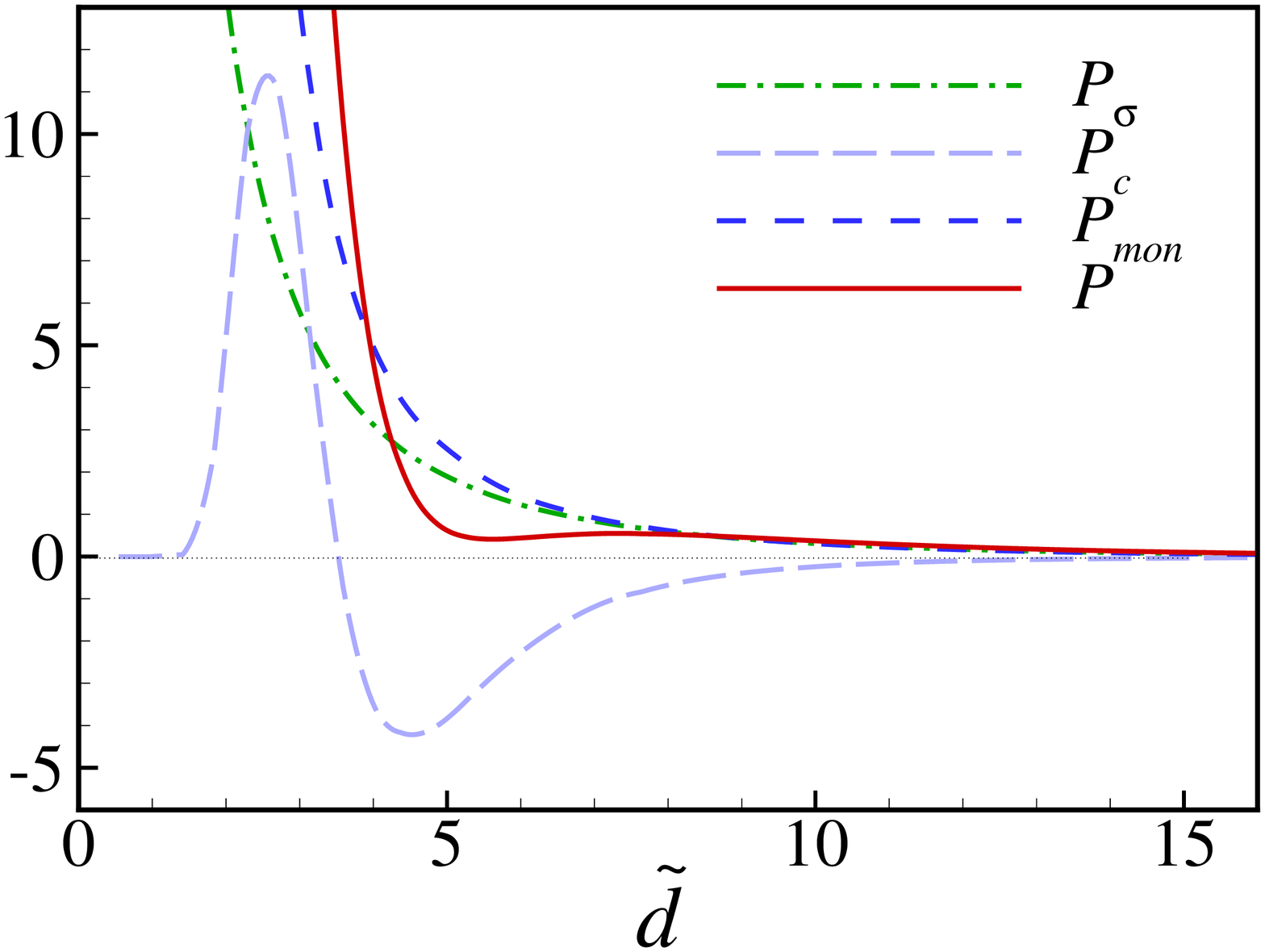} (b)
	\end{center}\end{minipage}
    \begin{minipage}[h]{0.32\textwidth}\begin{center}
		 \includegraphics[width=\textwidth]{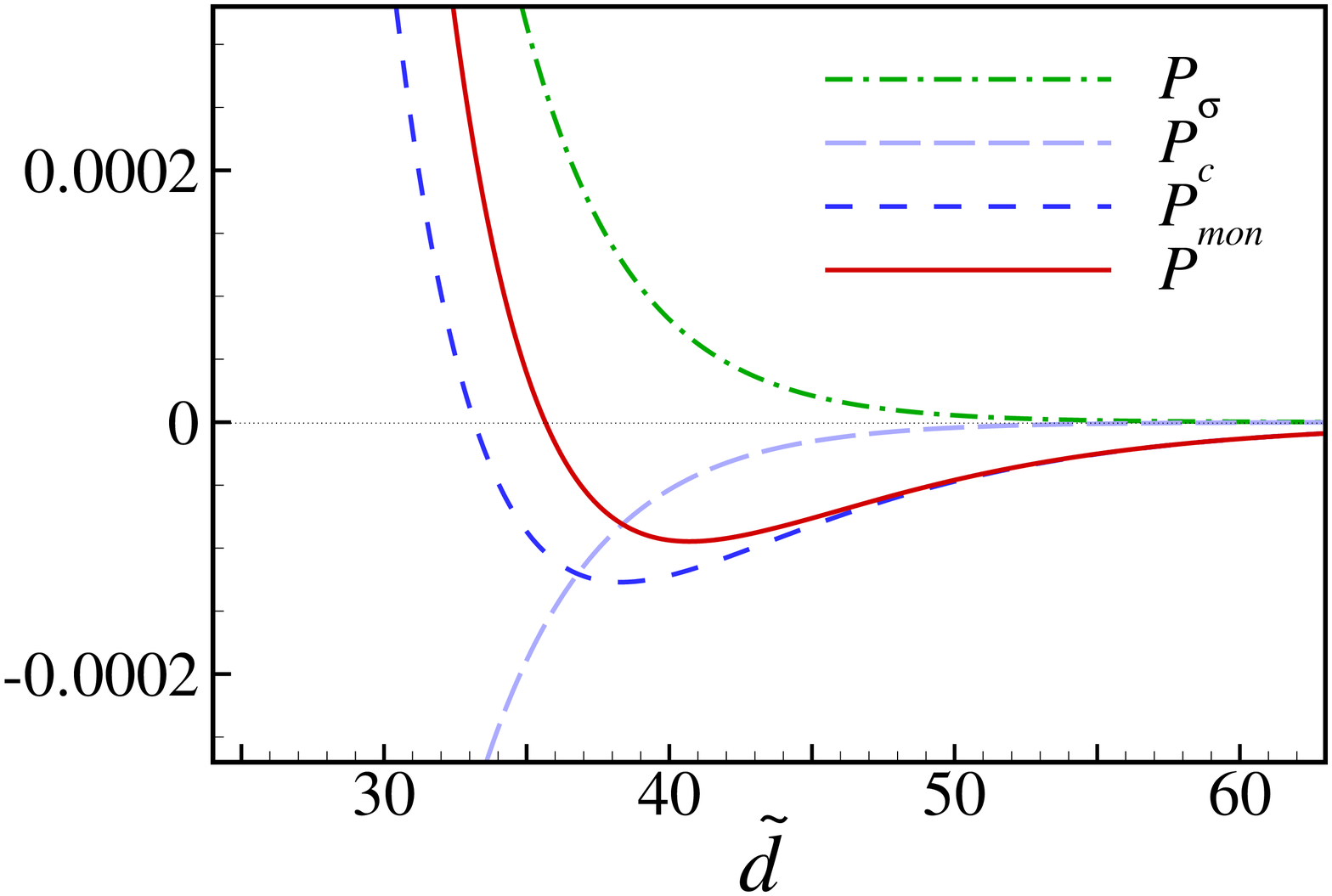} (c)
	\end{center}\end{minipage}
\caption{(Color online) (a) Rescaled interaction pressure  as a function of the rescaled distance between the uniformly charged inner surfaces of two identical semi-infinite dielectric slabs  with $\Xi=50$, $\Delta=0.95$ and $\tilde \kappa=0.27$ and for a few different values of the rescaled multivalent counterions concentration, $\tilde \chi_c$, as indicated on the graph. Inset shows a close-up view of the region around the shallow local minimum at large separations. Panels (b) and (c) show different components that contribute to the interaction pressure for $\tilde \chi_c=0.11$ in the vicinity of the small-separation and large-separation minima, respectively  (see the text for definitions). }
\label{fig:pressure_chic_no disorder}
\end{center}\end{figure*}

\subsection{Interaction of non-disordered surfaces}
\label{subsec:non-dis}

We first consider the interaction pressure between two non-disordered (uniformly charged) surfaces within the dressed multivalent-ion theory.  The interaction pressure acting on the slabs in this case follows from Eqs. (\ref{eq:P_total})-(\ref{eq:P_mon}) by noting that in all the expressions involved we need to set $\chi=0$; hence, in particular we have $\tilde P_{dis}=0$. We shall primarily focus on the case of semi-infinite slabs ($b=\infty$) and fix the electrostatic coupling parameter at   $\Xi=50$, which can be achieved with tetravalent counterions ($q=4$) and the mean surface charge density  $\sigma=0.24~{\mathrm{nm}}^{-2}$ in water ($\epsilon_m=80$) and at room temperature $T=293$~K (see Table \ref{table}). Unless otherwise specified, we take $\Delta=0.95$, which is appropriate for water/hydrocarbon interfaces.

The results for the rescaled interaction pressure  are shown in Fig. \ref{fig:pressure_no disorder}a as a function of the rescaled inter-surface separation for a few different values of the rescaled screening parameter, i.e., $\tilde \kappa=0.25, 0.27, 0.28, 0.29, 0.30$, and  $\tilde \chi_c=0.15$ (in actual units, these parameter values can be obtained, for instance, by taking  salt bulk concentrations $n_0 \simeq105, 123, 133, 143, 153$~mM and  $c_0\simeq 2.5$~mM).
At sufficiently large $\tilde \kappa$ (e.g., high monovalent salt concentration), the pressure is repulsive and decays monotonically with the distance. In this regime, the salt screening effects are dominant and the SC effects due to the multivalent counterions are strongly suppressed; the repulsive pressure is in fact determined mainly by the surface-surface repulsion ($\tilde P_\sigma$) and the osmotic pressure of monovalent ions ($\tilde P_{mon}$) with the latter contribution being the larger of the two, as can be seen from the pressure components in Fig. \ref{fig:pressure_no disorder}b.

As $\tilde \kappa$ is decreased, the pressure develops a non-monotonic behavior with a pronounced local minimum at intermediate inter-surface separations within a range comparable to the Debye screening length (Fig. \ref{fig:pressure_no disorder}a). This local minimum eventually turns into a negative global minimum as $\tilde \kappa$ is decreased further, indicating a strong attractive pressure induced between the slabs. For $\tilde \kappa=0.27$ (red dashed curve), the minimum attractive pressure is approximately $\tilde P\simeq -2.7$
(or, with the choice of the actual parameter values mentioned above, $P\simeq -28$~bar).
For $\tilde \kappa=0.25$ (black solid curve in Fig. \ref{fig:pressure_no disorder}a which is replotted as the red solid curve in Fig. \ref{fig:pressure_no disorder}c), the minimum attractive pressure is $\tilde P\simeq -75$
(or, $P\simeq -780$~bar).
Note that the net attraction here appears despite the fact that the surfaces are {\em like charged}. This is because of the SC effects, which are produced by the leading surface-counterion correlations   \cite{hoda_review, perspective, Naji_PhysicaA,Netz01,AndrePRL,AndreEPJE} and enter through the pressure component  $\tilde P_c$. This contribution becomes  quite large  at  small screening parameters (see Fig. \ref{fig:pressure_no disorder}c) and exhibits a non-monotonic behavior  that we shall consider later in more detail. The non-monotonic behavior of the net pressure with the inter-surface separation (Fig. \ref{fig:pressure_no disorder}a) stems directly from the interplay between its different components, with $\tilde P_c$ being the most essential one.

Similar behavior to those shown in Fig. \ref{fig:pressure_no disorder}a can be seen upon increasing the rescaled bulk concentration of multivalent counterions, $\tilde \chi_c$ (Fig. \ref{fig:pressure_chic_no disorder}), and upon decreasing the dielectric discontinuity parameter, $\Delta$ (not shown) with the minimum attractive pressure turing out to be quite sensitive to the exact values of these parameters. Figure \ref{fig:pressure_chic_no disorder}a shows the results for $\tilde \chi_c=0.09, 0.11, 0.13, 0.15$ and $\tilde \kappa=0.27$  (corresponding, for instance, to choosing the bulk concentrations as $c_0\simeq 0.9, 1.3, 1.9, 2.5$~mM  with  $n_0 \simeq 127, 126, 125, 123$~mM, respectively). 
The pressure still remains strongly repulsive at very small separations because of the contributions $\tilde P_\sigma$ and $\tilde P_{mon}$ discussed above (see Fig. \ref{fig:pressure_chic_no disorder}b and also Figs. \ref{fig:pressure_no disorder}b and c). However, the multivalent pressure component  $\tilde P_c$ again shows a non-monotonic behavior with both attractive and repulsive regions (Fig. \ref{fig:pressure_chic_no disorder}b).

\begin{figure}[t!]
\begin{center}
     \begin{minipage}[h]{0.32\textwidth}\begin{center}
		\includegraphics[width=\textwidth]{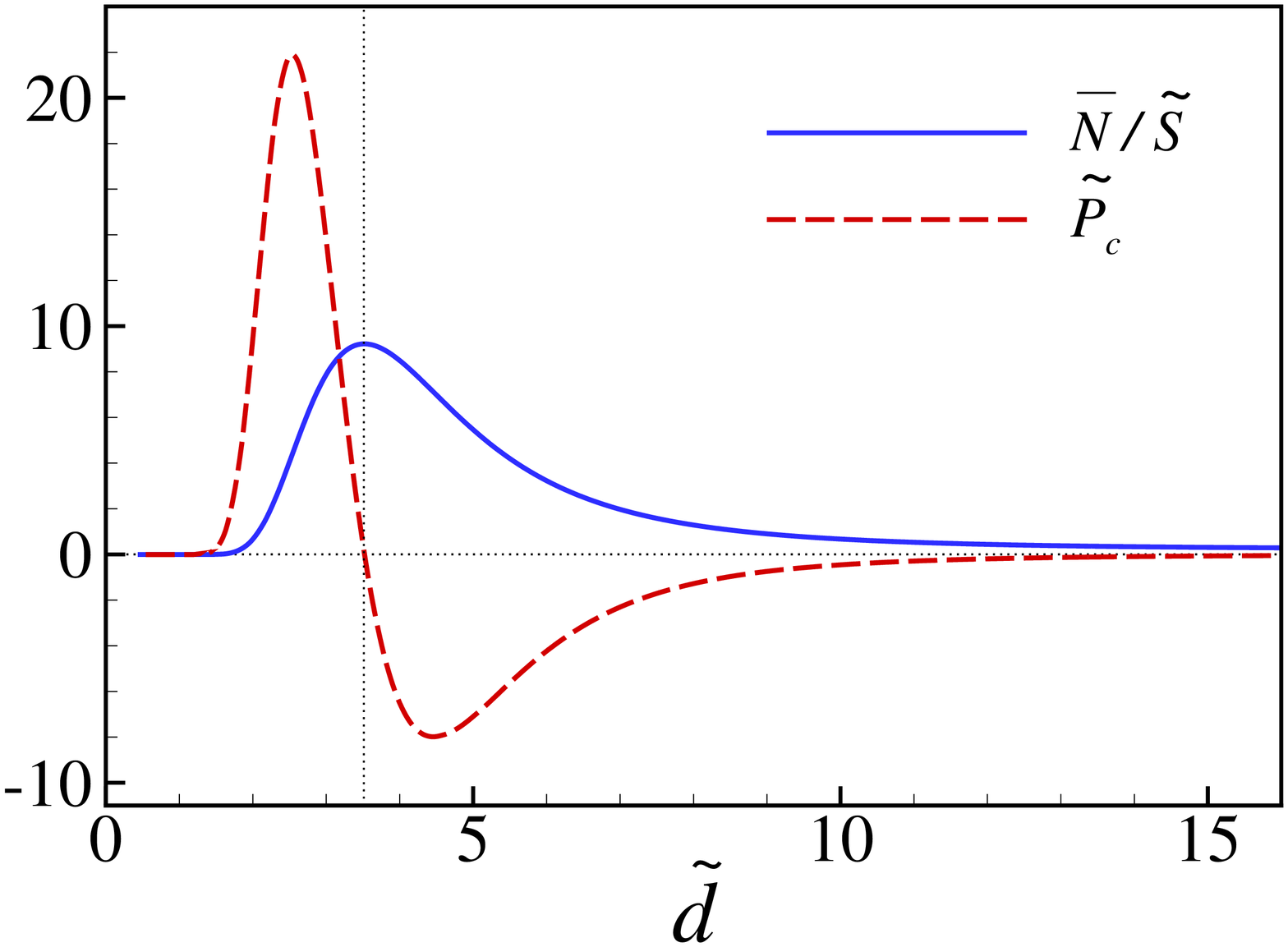}
	\end{center}\end{minipage}	
\caption{Rescaled average number of multivalent counterions  in the slit per rescaled surface area (blue curve) and the corresponding rescaled pressure component, $\tilde P_c$ (red dashed curve), as a function of the rescaled distance between the uniformly charged inner surfaces of two identical semi-infinite dielectric slabs  for $\tilde \kappa=0.27$, $\tilde \chi_c=0.15$, $\Xi=50$ and $\Delta=0.95$.}
\label{fig:NP}
\end{center}
\end{figure}

The non-monotonic behavior of $\tilde P_c$ can be understood by inspecting the average number of multivalent counterions in the slit between the slabs, $\bar{N}= S  \int_{-d/2}^{d/2}\rmd z\, c(z)$, which, in rescaled units and per unit area, is given by
\begin{equation} \label{eq:Dis_neutral}
\frac{\bar{N}}{\tilde S}=\frac{\tilde \chi_c^2}{8\pi\Xi} \int_{-\tilde d/2}^{\tilde d/2}\rmd \tilde z\, \tilde c(\tilde z).
\end{equation}
This quantity is shown in Fig. \ref{fig:NP} for $\tilde \kappa=0.27$ and $\tilde \chi_c=0.15$  along with its corresponding pressure component $\tilde P_c$ that in fact correspond to the red dashed curve in Fig. \ref{fig:pressure_no disorder}a). At large separations and upon decreasing the inter-surface distance, the number of multivalent counterions in slit is increased due to a larger uptake of these ions from the bulk solution, which is caused by an increased counterion-surface correlation (attraction) that, in turn, enhances the attractive (negative) pressure component  $\tilde P_c$. As the separation is decreased further, the number of multivalent counterions in the slit reaches a maximum value. The multivalent counterions at smaller inter-surface separations are strongly repelled by their image charges, are depleted from the slit and eventually completely ejected from the slit when the surfaces come close to contact. The pressure component due to multivalent counterions thus changes sign at the location where $\bar N$ reaches a maximum and, eventually, tends to the bulk pressure $\tilde P_c\rightarrow -\tilde \chi_c^2/4$ (or, in actual units, $P_c\rightarrow -c_0 k_B T$) when $\tilde d\rightarrow 0$ (this limiting value is not discernible at the range of scales shown in the figure).

\begin{figure*}[t!]\begin{center}
	\begin{minipage}[h]{0.32\textwidth}\begin{center}
		\includegraphics[width=\textwidth]{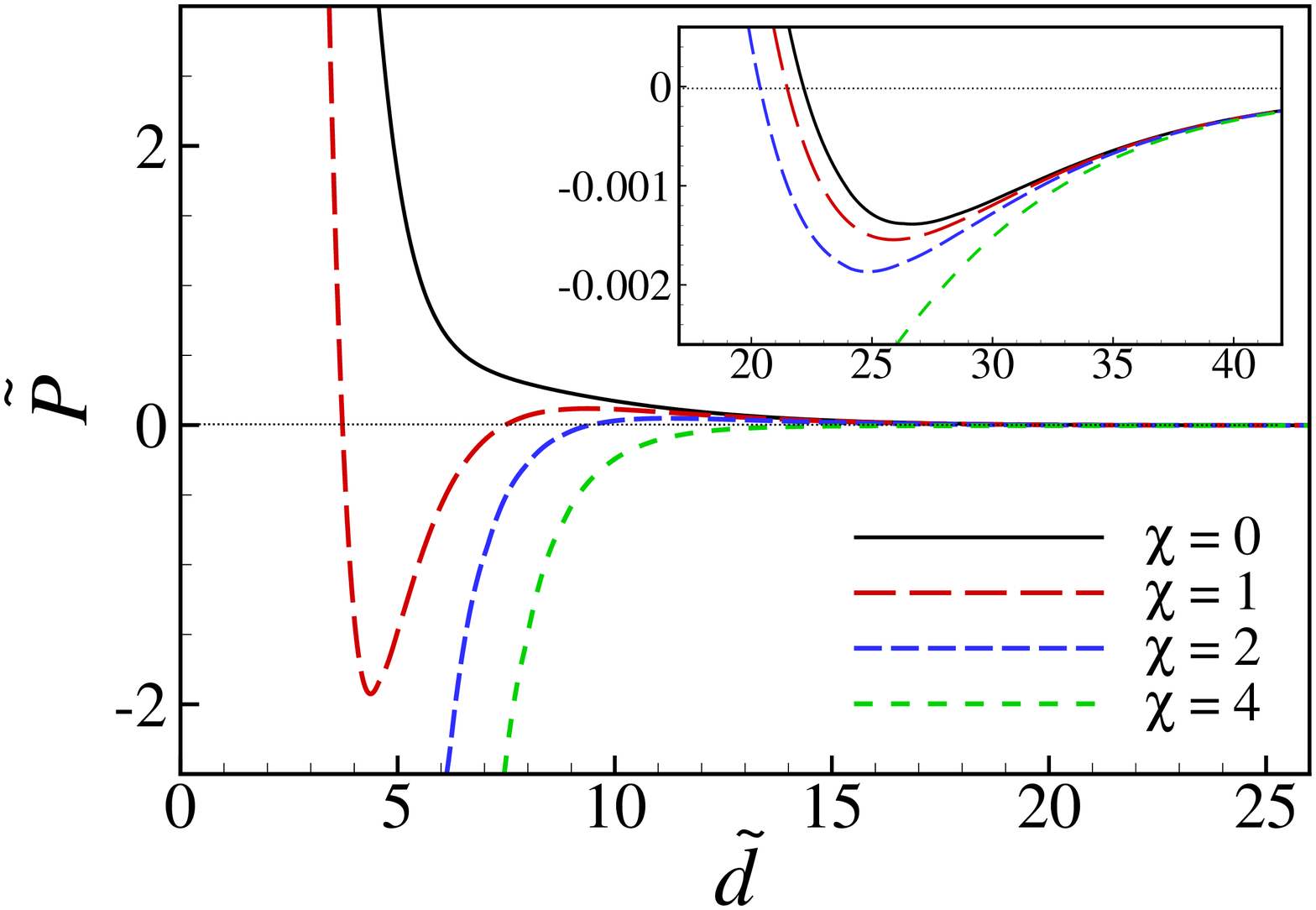} (a)
	\end{center}\end{minipage}\hskip -0.05cm	
    \begin{minipage}[h]{0.32\textwidth}\begin{center}
		\includegraphics[width=\textwidth]{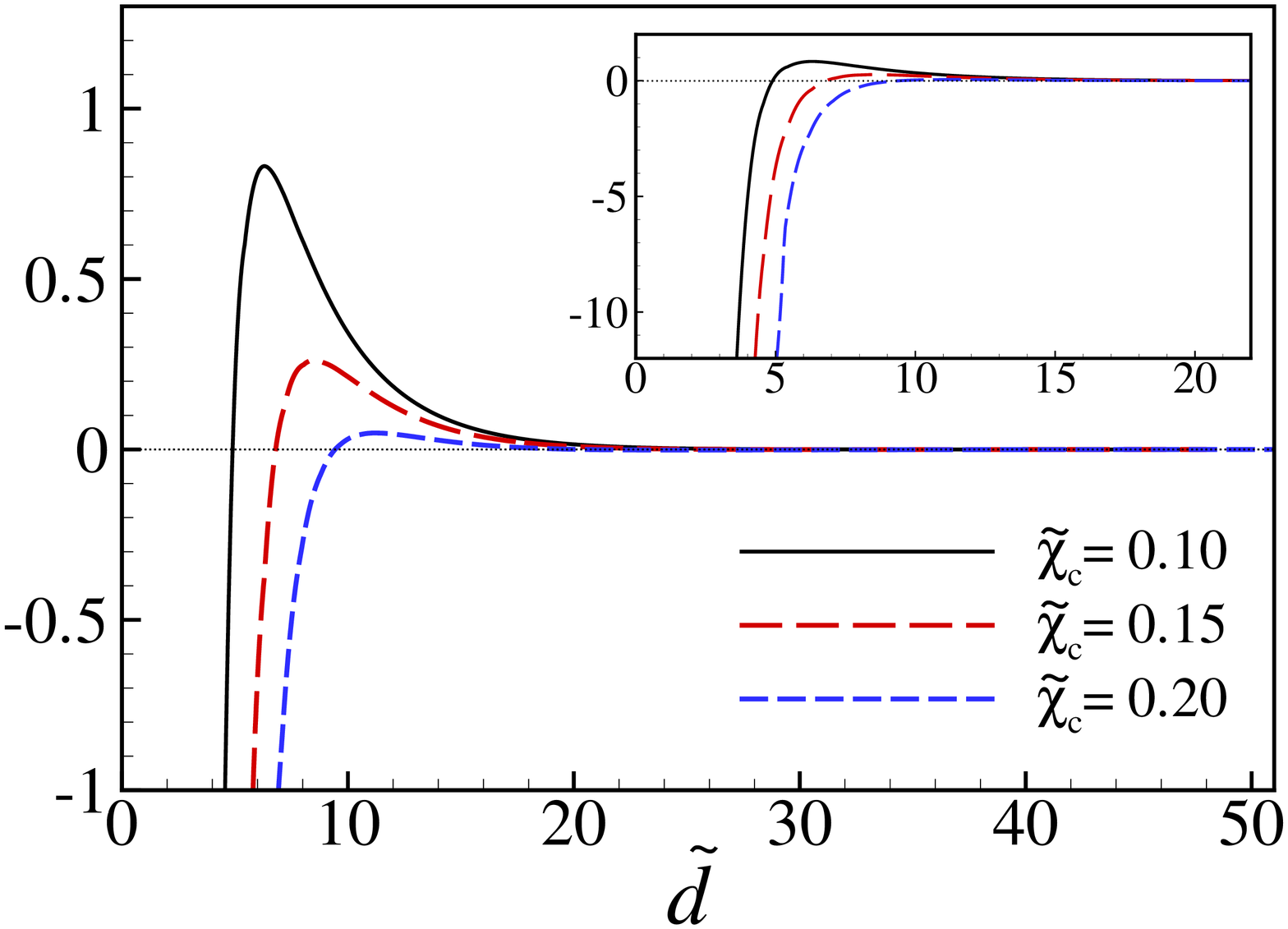} (b)
	\end{center}\end{minipage}\hskip -0.05cm	
	\begin{minipage}[h]{0.32\textwidth}\begin{center}
		 \includegraphics[width=\textwidth]{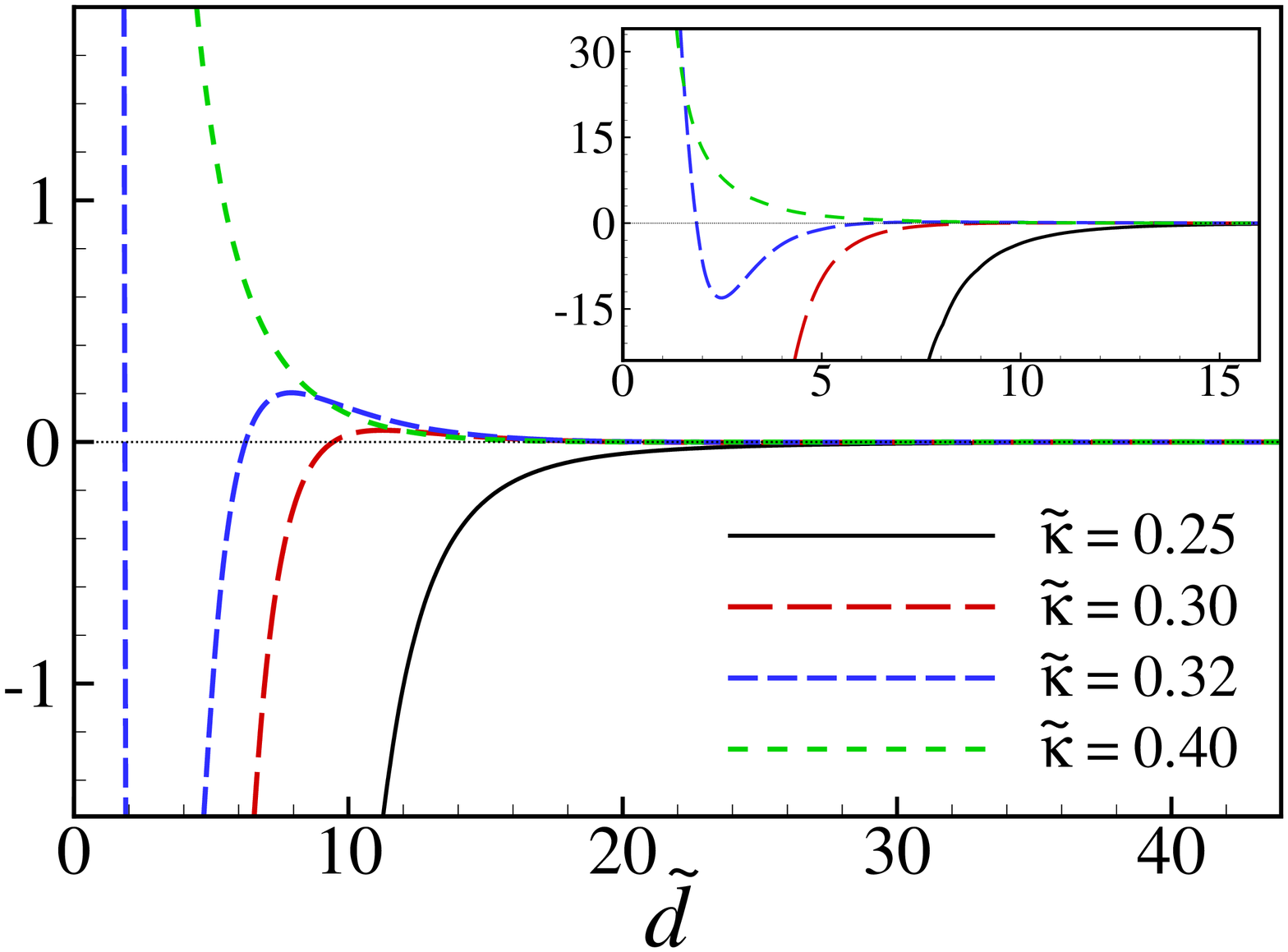} (c)
	\end{center}\end{minipage}
\caption{(Color online) Rescaled interaction pressure  as a function of the rescaled distance between the randomly charged inner surfaces of two semi-infinite dielectric slabs  with fixed   $\Xi=50$ and $\Delta=0.95$ and for (a) fixed $\tilde \kappa=0.3$, $\tilde\chi_c =0.2$ and different values of $\chi$, (b)  fixed $\tilde \kappa=0.3$, $\chi=2$ and different values of $\tilde\chi_c$, and (c) fixed $\tilde\chi_c =0.2$, $\chi=2$ and different values of $\tilde \kappa$ as shown on the graphs. Inset in  (a) shows a close-up view of the region around the shallow local minimum at large separations. Insets in  (b) and (c) show the behavior of the rescaled interaction  pressure over a wider range of scales.
}
\label{fig:inf_pressure}
\end{center}\end{figure*}

We should also note that the intermediate attractive-pressure regime seen in Figs. \ref{fig:pressure_no disorder}a and \ref{fig:pressure_chic_no disorder}a is followed by a weakly repulsive regime at larger separations but there is a very shallow local minimum at large separations that gives the pressure curves a weakly attractive long tail. This is  shown  in a close-up view in the inset of Fig. \ref{fig:pressure_chic_no disorder}a for the curves that appear in the main set. Comparing the different pressure components  around this large-separation minimum (see Fig. \ref{fig:pressure_chic_no disorder}c) shows that  the contribution from the surface-surface repulsion ($\tilde P_\sigma$) nearly cancels the attractive contribution from multivalent counterions  ($\tilde P_c$)  and, thus, the total pressure is determined almost completely by the contribution from monovalent ions ($\tilde P_{mon}$), which itself shows a shallow minimum at large separations.

In general, the depth of this minimum, and the large-separation attraction regime, can be enhanced slightly also by $\tilde P_c$ (not shown). A closer inspection of $\tilde P_{mon}$ shows that the partial osmotic pressure that contributes to this component from monovalent salt anions is negative (because these ions have the same sign as the surface charges and are therefore depleted from the slit) and is slightly larger in magnitude than the (positive) partial osmotic pressure from monovalent salt cations, hence, causing the weakly attractive, large-separation behavior of $\tilde P_{mon}$. This behavior is robust and appears for a wide range of parameter values.

Another important point to be considered here is whether the van-der-Waals-like \cite{VDW} non-monotonicity observed in the pressure-distance curves  in Figs. \ref{fig:pressure_no disorder}a  and \ref{fig:pressure_chic_no disorder}a gives any indication of a phase coexistence, if the pressure curves are to be interpreted in the  thermodynamic sense as, e.g., the pressure in a stack of plane-parallel like-charged membranes? Such a phase coexistence would imply a simultaneous existence of a dense phase and a swollen phase in the system. However, if one applies the Maxwell equal-area construction, the negative area under the pressure curves, which comes from the region around the attractive minimum at small separations,  is typically much larger than the  positive area coming from the region around the repulsive maximum (hump) at intermediate separations. This translates into a statement that the equal-area Maxwell construction cannot be fulfilled in general.

\subsection{Interaction of disordered surfaces}
\label{subsec:dis}

We now consider the situation where the inner surfaces of the slabs carry a finite degree of charge disorder characterized by the dimensionless coupling parameter, $\chi$. The rescaled interaction pressure again follows from Eqs. (\ref{eq:P_total})-(\ref{eq:P_mon}) and its behavior as a function of the rescaled inter-surface separation is shown in Fig. \ref{fig:inf_pressure} for a few different sets of parameters in the case of semi-infinite slabs ($b=\infty$) with fixed  $\Xi=50$ and  $\Delta=0.95$.

In Fig. \ref{fig:inf_pressure}a, we show the results for  a few different values of the  disorder coupling parameter $\chi=0, 1, 2$ and 4 with $\tilde \kappa=0.3$ and $\tilde\chi_c =0.2$. With the choice of physical parameter values as $q=4$ (tetravalent counterions),  $\sigma=0.24~{\mathrm{nm}}^{-2}$ for water  ($\epsilon_m=80$)  at room temperature ($T=293$~K), these rescaled parameter values  correspond to surface charge disorder variances  $g=0, 0.02, 0.04$ and  $0.08~{\mathrm{nm}}^{-2}$, respectively, and mono- and multivalent salt bulk concentrations of $n_0\simeq 150$~mM and $c_0\simeq 4.4$~mM  (see  Table \ref{table}).

For  the non-disordered case with $\chi=0$  (black solid curve), the results show a  repulsive, monotonically decaying interaction at small separations and a shallow attractive minimum at larger separations around  $\tilde d \simeq 25$ of the type that were analyzed in detail in Figs. \ref{fig:pressure_no disorder} and \ref{fig:pressure_chic_no disorder}. The presence of charge disorder on the inner surfaces of the slabs leads to significant qualitative differences in the effective interaction profile of the two surfaces (dashed curves). The disorder effects dominate at small to intermediate separations and turn the surface repulsion to a very strong attractive interaction. They diminish at larger separations as the curves with different values of $\chi$ (including $\chi=0$) converge. These features indicate an interplay between different contributions to the net interaction  pressure that we shall examine later.

The attractive regime at small separations is followed by a repulsive  regime with a pronounced hump at intermediate separations.  For the set of parameter values used in Fig. \ref{fig:inf_pressure}a and for $\chi =2$ (blue dashed curve), the position of the hump and the maximum pressure are given by $\tilde d\simeq 11.2$ and  $\tilde P\simeq 0.05$, which, for the set of the physical parameter values mentioned above, correspond to $d\simeq 2.6$~nm and $P\simeq 0.5$~bar. The shallow attractive minimum of the pressure curves at larger separations ($\tilde d\simeq 25$ or $d\simeq 5.7$~nm) is only weakly influenced by the presence of disorder, as expected.

For sufficiently large disorder strengths (e.g., for $\chi =4$ shown by the green dashed curve), the attractive regime  extends to the whole range of small to intermediate separations, even beyond the large-separation minimum, as shown in the inset, producing thus a long-ranged attractive interaction between the slabs. This  follows as a consequence of the combined effects from the SC electrostatics of multivalent counterion and the surface charge disorder that  couple to one another through the multivalent counterions contribution, $\tilde P_c$ (see below), leading to features that are distinctly different from what we found in the case of uniformly charged surfaces (Section \ref{subsec:non-dis}).

A similar trend is found if the disorder strength is kept fixed and the rescaled bulk concentration of multivalent counterions, $\tilde \chi_c$,  is increased (see Fig. \ref{fig:inf_pressure}b). In this case both the range and the  strength of the attractive pressure acting on the slabs is increased. We find a slightly different behavior when the Debye screening parameter (or equivalently, the bulk concentration of monovalent salt) is changed. For the parameters shown in Fig. \ref{fig:inf_pressure}c, we see that, as $\tilde \kappa$ is decreased, the interaction pressure at  intermediate separations first increases and becomes slightly more repulsive (compare green and blue dashed curves) and then turns to become attractive. This is in accord with the intuitive expectation that the SC  and disorder-induced effects become  stronger at lower salt concentrations.

\begin{figure}
     \begin{minipage}[h]{0.32\textwidth}\begin{center}
		\includegraphics[width=\textwidth]{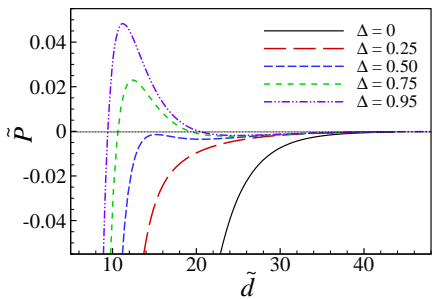}
	 \end{center}\end{minipage}
  \caption{(Color online) Same as Fig. \ref{fig:inf_pressure} but for fixed  $\Xi=50$, $\tilde \kappa=0.3$, $\tilde\chi_c =0.2$, $\chi=2$ and different values of $\Delta$ as shown on the graph.}
  \label{fig:Pdelta}
\end{figure}

The disorder-induced attraction is enhanced and the repulsive hump and the large-separation minimum in the pressure curves disappear also when 
$\Delta \rightarrow 0$ (see Fig. \ref{fig:Pdelta}), which clearly points to the key role of dielectric images, generating stronger  repulsive interaction in systems with larger interfacial dielectric discontinuity. In the dielectrically homogeneous case (i.e., with $\Delta=0$ and $\tilde \kappa>0$), we find long-ranged,  monotonic attraction in the whole range of inter-surface separations with a diverging attractive pressure in the limit  $\tilde d\rightarrow 0$ for all values of $\chi\geq 0$ (note   that this behavior is in contrast with what we found in the canonical counterion-only case in Section \ref{subsec:ci-only}, where the pressure for $\tilde d\rightarrow 0$ becomes repulsive in the regime $\chi<1$). When the dielectric image effects are included ($\Delta>0$), the pressure turns repulsive again at small separations as, e.g., seen for $\chi =1$ in Fig. \ref{fig:inf_pressure}a and for $\tilde \kappa = 0.32$ in Fig. \ref{fig:inf_pressure}c; the other curves with larger $\chi$ or smaller $\tilde \kappa$ in these figures, and also those shown in Fig. \ref{fig:inf_pressure}b, turn repulsive at extremely small separations, where the counterion-image repulsions eventually dominate and the multivalent counterions are fully depleted from the slit region (see also the discussion relating to Figs. \ref{fig:inf_dist}b in Section \ref{subsec:Density_salt}). However, the upturn of the pressure curves in these latter cases occurs at separations that are not physically meaningful  (insofar as the experimental realizations of our model are concerned), e.g., around $\tilde d\simeq 0.02$ (or $d\simeq 0.005$~nm)
for $\chi=2$, the blue dashed curve, in Fig. \ref{fig:inf_pressure}a, and, thus, for the sake of presentation, it has not been shown in the graphs. This, on the other hand,  means that, for sufficiently large $\chi$ and $\tilde \chi_c$ and/or sufficiently small $\tilde \kappa$, the stable equilibrium separation between the slabs (corresponding to the smallest point of zero pressure as, e.g., seen in Figs. \ref{fig:pressure_no disorder}a and \ref{fig:pressure_chic_no disorder}a in the case of non-disordered surfaces and in Figs. \ref{fig:inf_pressure}a and c in the case of disordered surfaces) is pushed down to very small values, where the surfaces are nearly in contact. Indeed, in the case of  disordered (but otherwise effectively like-charged) surfaces with no interfacial dielectric discontinuity and with  counterions only,  the disorder-induced attraction can  become strong enough to cause a continuous  transition to a collapsed state  \cite{ali-rudi,partial}. 

\begin{figure}[t!]
  	\begin{minipage}[h]{0.32\textwidth}\begin{center}
		\includegraphics[width=\textwidth]{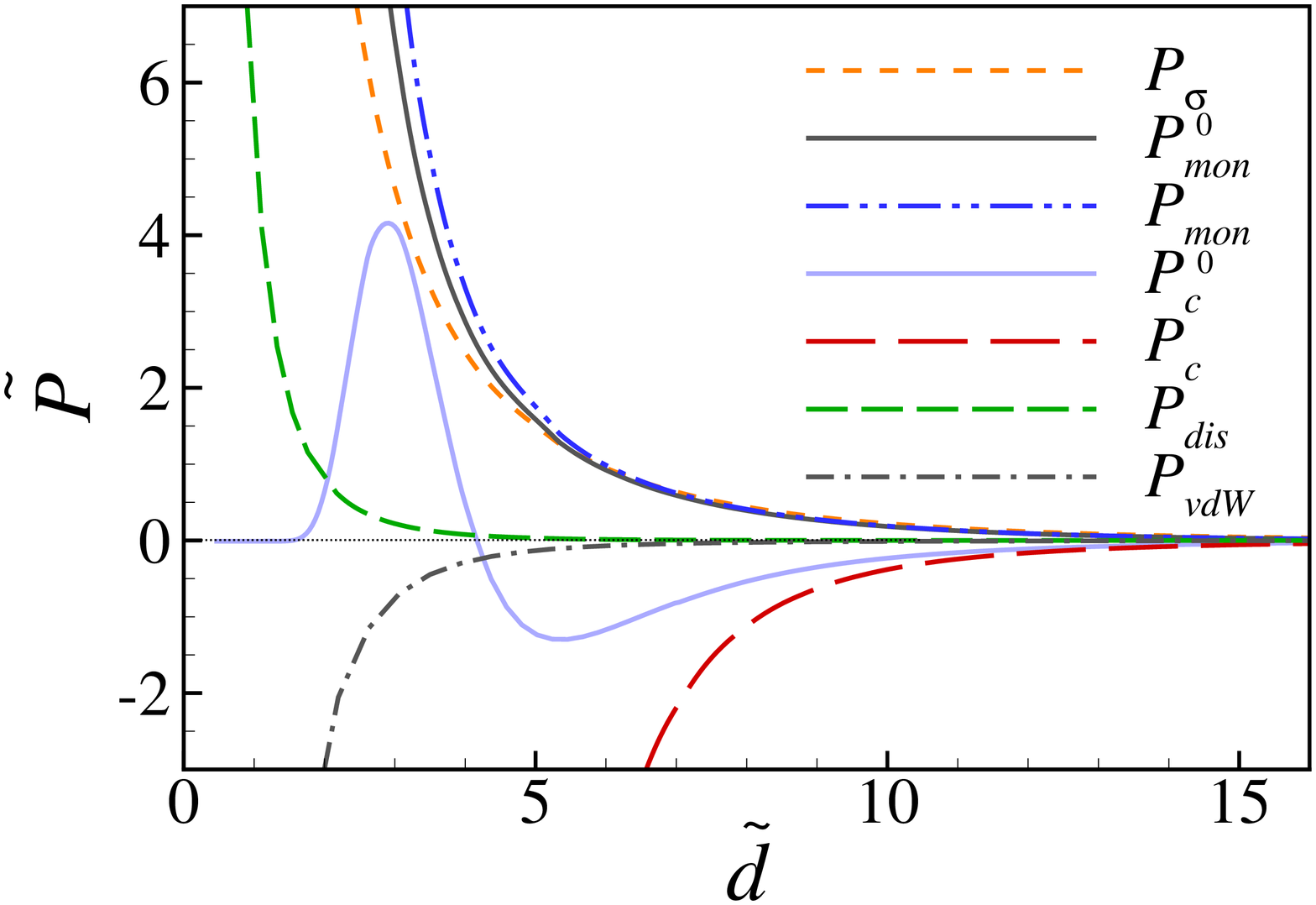} (a)
	\end{center}\end{minipage}\vskip 0.2cm	
    \begin{minipage}[h]{0.32\textwidth}\begin{center}
		\includegraphics[width=\textwidth]{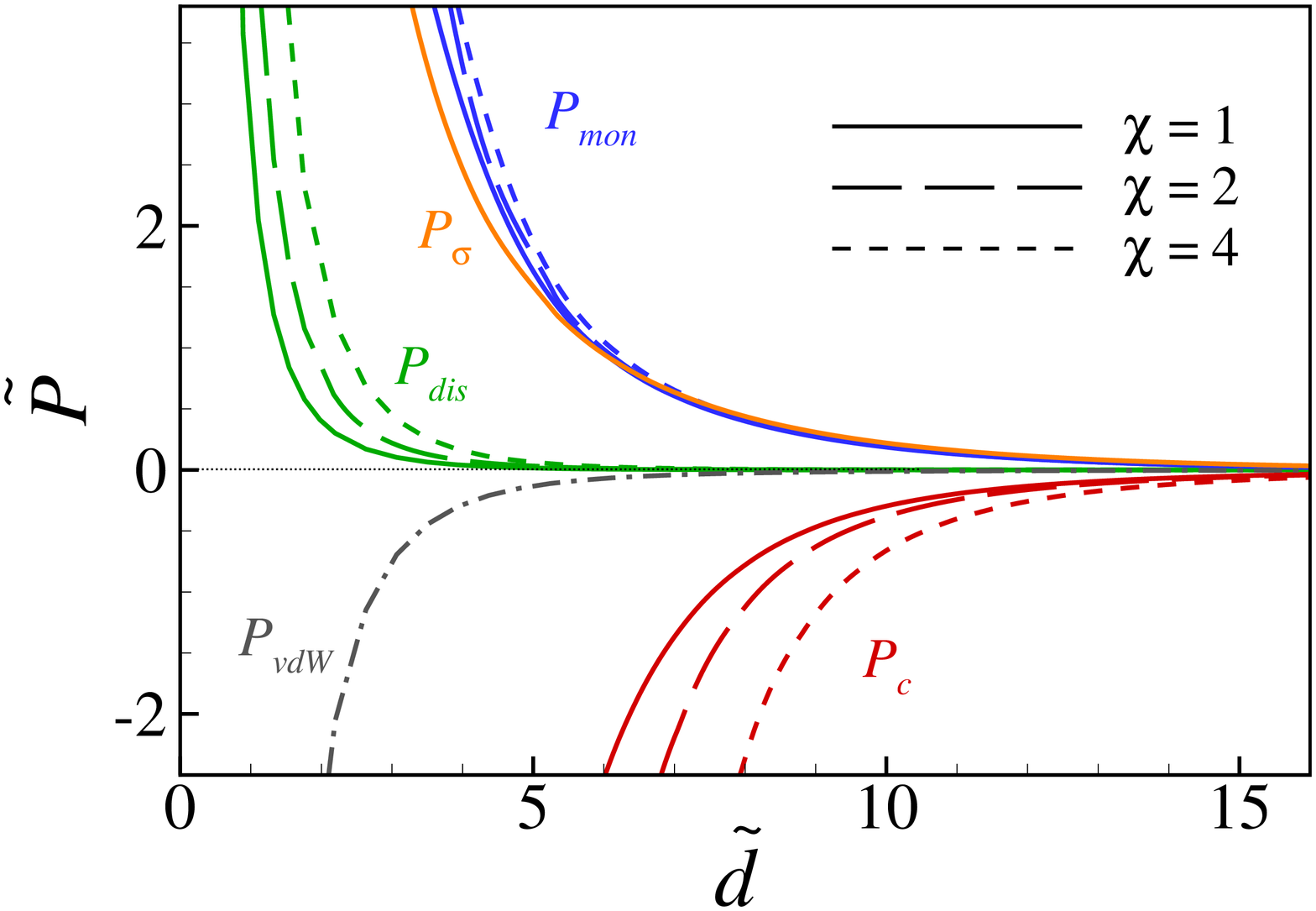} (b)
	\end{center}\end{minipage}
  \caption{(Color online) (a) Different components that contribute to the rescaled interaction pressure plotted as a function of the rescaled inter-surface distance between the slabs for the cases with disordered ($\chi=2$) and non-disordered (uniformly charged, $\chi=0$) surfaces. The curves that correspond to the non-disordered case are marked in the legends with the superscript `0'. The system parameters are $\Xi=50$, $\Delta=0.95$, $\tilde \kappa=0.3$ and $\tilde\chi_c =0.2$. For comparison, we also show the vdW pressure between the slabs.  (b) Different rescaled pressure components for the disordered case for the same parameter values as in (a) but with $\chi=1, 2$ and 4. }
  \label{fig:Pcomp}
\end{figure}

The disorder-induced non-monotonicity found in the interaction profiles can be understood by examining the different components that contribute to the net pressure as defined in Eqs. (\ref{eq:P_total})-(\ref{eq:P_mon}). These are the mean surface-surface repulsion, $\tilde P_{\sigma}$,  the interaction pressure due to the disorder variance, $\tilde P_{dis}$,  the interaction pressure mediated by the multivalent counterions, $\tilde P_{c}$, and the  monovalent ions contribution, $\tilde P_{mon}$. The effects of charge disorder, the dielectric and salt images   are systematically included in all these components.  In Fig. \ref{fig:Pcomp}a, we show their behavior as a function of the rescaled inter-surface separation for  $\Xi=50$, $\Delta=0.95$, $ \tilde \kappa=0.3$,  $\tilde \chi_c=0.2$ and $\chi=2$. In order to enable a comparison between the non-disordered and disordered cases, we also show the corresponding curves of the non-disordered case with $\chi=0$ (marked in the legends with the superscript `0'). Note that $\tilde P_{\sigma}$ is independent of the disorder strength and decays monotonically with the inter-surface distance. This  contribution is comparable to that from the monovalent ions  $\tilde P_{mon}$, which is also repulsive and shows a very weak dependence on the disorder strength with a  slightly larger (more repulsive) value in the case of disordered surfaces as compared with non-disordered ones ($\tilde P_{mon}>\tilde P_{mon}^0$). This is because, by increasing the disorder strength, more monovalent ions are attracted to the slit from the bulk, creating a larger osmotic pressure component. The effects of surface charge disorder on the multivalent counterions pressure component are quite substantial (compare $\tilde P_c$ and $\tilde P_c^0$) and most significant at small to intermediate separations, where the non-disordered contribution, $\tilde P_c^0$ (light-blue solid curve)  shows a highly non-monotonic behavior with a repulsive (positive)  hump as discussed  in Section \ref{subsec:non-dis}, while the disorder contribution,  $\tilde P_c$ (red dashed curve),  is attractive (negative)  and  increases in strength monotonically as the separation distance becomes small. Again, the disorder effects diminish and these two cases converge at separations larger than the Debye screening length.

The other  pressure component  that needs to be considered here is $\tilde P_{dis}$ that stems from the disorder variance and exists irrespective of the multivalent counterions in the system. This contribution is repulsive for $\Delta>0$  and becomes relevant only at very small separations. For the sake of comparison, we also show the inter-surface pressure generated by the van-der-Waals (vdW) interaction between two semi-infinite slabs immersed in an ionic mixture (dot-dashed curve). This interaction pressure can be calculated from the standard Lifshitz theory as (in actual units) \cite{Parsegian2005}
\begin{equation}
\label{p:vdwz}
P_{vdW}= - k_{\mathrm{B}}T\int_0^\infty \frac{Q \rmd Q}{2\pi }\frac{\gamma \Delta_s^2 \,\rme^{-2 \gamma d}}{1 - \Delta_s^2\, \rme^{-2 \gamma d}} -\frac{A }{6\pi d^3},
\end{equation}
where the first term comes from the zero-frequency mode of the electromagnetic field-fluctuations and the second term comes from the higher-order Matsubara frequencies. 
$A$ is the so-called Hamaker coefficient, which we take as $A=3\, \text{zJ}$ (upper bound for the non-zero Matsubara modes in the case of hydrocarbon slabs interacting across an aqueous medium \cite{hamakar}).  For slabs of lower dielectric polarizability than the solvent, such as the cases considered here,  the vdW interaction is attractive. In the figures, we show the rescaled quantity $\tilde P_{vdW} = \beta P_{vdW}/(2\pi \ell_{\mathrm{B}} \sigma^2)$ for the given parameter values. It is clear from Fig. \ref{fig:Pcomp}a that the vdW component has a comparable range and strength as $\tilde P_{dis}$; these two contributions can thus compete at very small separations, and in the absence of multivalent counterions, generate a non-monotonic interaction profile between randomly charged dielectrics considered in Refs.  \cite{rudiali,disorder-PRL,jcp2010,pre2011,epje2012,jcp2012,preprint3}.
These effects are however masked by the multivalent counterions contribution, $\tilde P_c$.

If $\chi$ is increased further (Fig. \ref{fig:Pcomp}b), the contribution of monovalent ions $\tilde P_{mon}$ changes only slightly (becoming more repulsive as noted before), but the effect of the increase of $\chi$ on both $\tilde P_{dis}$ (which becomes more repulsive) and $\tilde P_c$ (which becomes more attractive) is rather substantial. Note that $\tilde P_{dis}$ depends linearly on the parameter $\chi$ according to Eq. (\ref{eq:P_dis}), while the dependence of  $\tilde P_c$ (and also $\tilde P_{mon}$) on $\chi$ is nontrivial and occurs through the  effective single-particle energy, Eq. (\ref{eq:ResU_neutral}).

\begin{figure}[t!]
  	\begin{minipage}[h]{0.32\textwidth}\begin{center}
		\includegraphics[width=\textwidth]{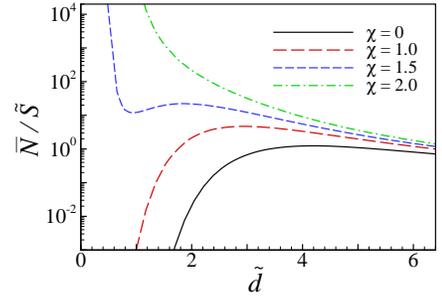}
	\end{center}\end{minipage}
  \caption{(Color online)
Rescaled average number of multivalent counterions  in the slit per rescaled surface area  as a function of the rescaled distance between the  randomly charged inner surfaces of two semi-infinite dielectric  slabs for a few different values of the disorder coupling parameter, $\chi$, for fixed  $\tilde \kappa=0.3$, $\tilde \chi_c=0.2$, $\Xi=50$ and $\Delta=0.95$.}
  \label{fig:N_dis}
\end{figure}

We can thus conclude that the qualitative differences found between the interaction pressure curves in the non-disordered and disordered systems are closely connected with the behavior of the multivalent counterions contribution, $\tilde P_c$, that becomes significantly more attractive and dominant at small to intermediate separations when  the disorder strength is increased, which explains the trends observed in Fig.  \ref{fig:inf_pressure}a). We should emphasize that the increase in the strength of this attractive component is due to the fact that a larger amount of multivalent counterions are pulled into the slit region because of the stronger counterion-surface attractions in the presence of disorder, mediating also a stronger inter-surface attraction. This can be seen by inspecting  the average number of multivalent counterions in the slit as shown in Fig. \ref{fig:N_dis}.

\begin{figure}[t!]
     \begin{minipage}[h]{0.32\textwidth}\begin{center}
		\includegraphics[width=\textwidth]{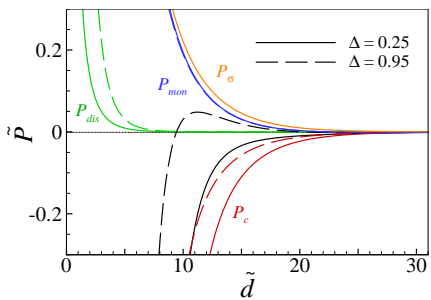}
	 \end{center}\end{minipage}
  \caption{(Color online) Same as Fig. \ref{fig:Pcomp}a but for two different values of $\Delta=0.25$ and 0.95 (the total pressure for these cases is shown by black solid and black dashed curves, respectively). }
  \label{fig:Pdelta_components}
\end{figure}

A similar behavior is found for $\tilde P_c$ when the rescaled bulk density of multivalent counterions is increased (Fig.  \ref{fig:inf_pressure}b), and/or when the Debye screening length (Fig.  \ref{fig:inf_pressure}c) or the dielectric discontinuity parameter (Fig. \ref{fig:Pdelta}) are decreased. For instance, we show the behavior of the different pressure components for $\Delta=0.25$ (solid curves)  and 0.95 (dashed curves) in Fig. \ref{fig:Pdelta_components}. Only the two disorder-induced components, $\tilde P_c$ and $\tilde P_{dis}$, show significant variations with the dielectric discontinuity parameter, $\Delta$. The changes in the interaction  pressure in Fig. \ref{fig:Pdelta} can be assigned primarily to the changes in the pressure component $\tilde P_c$ as, e.g., at larger values of $\Delta$, these ions are affected more strongly by  image repulsions and are depleted more strongly  from the slit region, giving a smaller attractive pressure, $\tilde P_c$.

\section{Conclusion and discussion}
\label{sec:discussion}

In this paper, we have studied the effective interactions mediated by multivalent counterions between dielectric slabs that carry a quenched, random monopolar charge distribution on their juxtaposed, plane-parallel, inner surfaces. We have considered both a case in which the Coulomb fluid in the slit between the slabs consists only of mobile multivalent counterions, and also a case in which the Coulomb fluid is an asymmetric ionic mixture containing a $q:1$ salt (with multivalent counterions of charge valency $q>1$) and a monovalent $1:1$ salt in equilibrium with a bulk reservoir. Our goal has been to elucidate the effects due to the coupling between the charge disorder  and electrostatic correlations in the so-called strong coupling (SC) regime, realized experimentally when the system contains mobile multivalent counterions, that give rise to strong surface-counterion correlations and also, to subleading orders, counterion-counterion correlations. The counter-intuitive phenomena in SC  electrostatics, such as attraction between like-charged surfaces, have been well-studied in the case of non-disordered (and, in most cases, uniformly charged) macromlecular surfaces  (see, e.g.,  recent reviews in Refs. \cite{hoda_review,perspective,Naji_PhysicaA,Levin02,Shklovs02} and references therein). The SC effects have, however, remained largely unexplored in the situation where the bounding (macromolecular) surfaces bear quenched, disordered charge distributions.

In the weak coupling (WC) regime, where, e.g., all ions in the Coulomb fluid are monovalent, the  quenched charge disorder on bounding surfaces turns out to have either no or only very small effects and does not lead to any qualitatively new features in the behavior of the Coulomb fluid \cite{ali-rudi,netz-disorder,netz-disorder2}. In the SC regime, the interaction of quenched random charge distributions across a Coulomb fluid has been considered only within the plane-parallel counterion-only model with no interfacial dielectric discontinuity \cite{ali-rudi,partial}. It was shown that quenched surface charge disorder leads to a renormalization (reduction) of the entropic contributions to the interaction free energy (corresponding to a renormalized effective temperature),  and that for sufficiently large disorder coupling parameter, i.e., $\chi\geq 1$, the surfaces undergo a continuous  transition to a collapsed state.

 In the present work, we have derived a generalized dressed multivalent-ion formalism that incorporates dielectric image effects as well as salt screening and salt image effects, being applicable to any arbitrary geometry of the bounding surfaces with disordered charge distributions. The charge disorder over the bounding surfaces is assumed to be Gaussian, without any spatial correlations, with possible generalizations to be discussed elsewhere \cite{preprint4}. The limiting case of this generalized model in the case of a single randomly charged dielectric slab has been discussed recently \cite{jcp2014}.

We have analyzed the prediction of the present theory in the specific example of two plane-parallel dielectric slabs with randomly charged inner surfaces immersed in an asymmetric Coulomb fluids. We have shown that, in a dielectrically homogeneous system (with or without additional salt), the multivalent counterions are attracted towards the surfaces with a singular, disorder-induced potential that diverges logarithmically in the vicinity of the surfaces; this creates a singular but integrable counterion density profile that exhibits an algebraic divergence at the surfaces with an exponent that depends on the disorder coupling parameter, $\chi$. Remarkably, this behavior is in contrast with the contact-value theorem, which describes the behavior of counterions at uniformly charged surfaces and predicts a finite contact density \cite{contact_value,contact_value2,contact_value1}.

\begin{table*}[t!]
    \begin{center}
\begin{tabular}{rrrrrrrr|rrrrrrr}
\hline\hline
$\tilde\kappa=0.2$\rule{0pt}{3ex}    &$0.25$    &    $0.3$ &   $0.4$&&&&& &$\chi$&&&$g\,({\mathrm{nm}}^{-2}$)&& \\
\hline
 \hspace{1ex} $n_0=68$\,mM   &\hspace{1ex} 108\,mM &\hspace{1ex} 156\,mM     &\hspace{1ex} 279\,mM     &\hspace{1ex}$c_0=$1.1\,mM    &\hspace{1ex} $\tilde\chi_c$= 0.1&&&&0.5 &&&0.01&&\\
  65\,mM  & 105\,mM &153\,mM & 277\,mM & 2.5\,mM &0.15& &&&1.0&&&0.02&&\\
 62\,mM & 101\,mM & 150\,mM    & 273\,mM     & 4.4\,mM& 0.20& &&&2.0&&&0.04&&\\
57\,mM  & 96\,mM &145\,mM & 268\,mM & 6.9\,mM &0.25& &&&4.0&&&0.08&&\\
\hline\hline
\end{tabular}
\caption{A few typical examples for the actual values of the bulk concentrations $n_0$ and $c_0$,  which can correspond to the typical values we have chosen for the rescaled parameters $\tilde\kappa$ and $\tilde\chi_c$. Here, we have fixed the other parameter values as $q=4$ (tetravalent counterions), $\ell_{\mathrm{B}}=0.71$~nm (corresponding to water with $\epsilon_m=80$ at room temperature, $T=293$~K) and surface charge density $\sigma=0.24~{\mathrm{nm}}^{-2}$, which give $\mu=0.23$~nm and $\Xi=50$. We also show the actual values of the disorder variance $g$, which can  correspond to a few typical  values of the disorder coupling parameter $\chi$  (see the text for definitions).
}
\label{table}
\end{center}
\end{table*}

Our results for the counterion-only case also shed further light on the previous findings for this system \cite{ali-rudi} that, as noted above, predicted a renormalized entropic contribution to the interaction free energy. We thus show here that such a behavior follows as a result of the singular accumulation of counterions in the immediate vicinity of the two surfaces, resulting in a renormalized temperature for the system. This notion of a renormalized temperature should be  used with caution because first, this particular form of the renormalization of the system entropy is found only in the SC limit of the two-surface counterion-only model, and secondly, one can show  that the renormalization in fact originates from both energetic and entropic sources. Nevertheless, the interplay between the translational entropy of strongly coupled counterions and the (non-thermal) configurational entropy, due to the averaging over different realizations of the quenched disorder, does result in the {\em anti-fragile} behavior \cite{taleb} of multivalent counterions. Antifragility does not only result in a more `ordered' state, characterized by a diminished intrinsic thermal `disorder' induced by the externally imposed disorder, but also in a thermodynamically more stable state since quenched surface charge disorder engenders also a lower free energy as compared with a non-disordered system. The appearance of antifragility is possibly one of the most fundamental and perplexing features in the complicated Coulomb world, whose ramifications we are only beginning to unravel.

The singular behavior of the multivalent counterion density on approach to randomly charged surfaces persists also when the dielectrically homogeneous system contains an additional monovalent salt component. The mobile monovalent salt ions generate screening effects at intermediate to large  distances  (comparable to the Debye screening length), while at small to intermediate distances from the bounding surfaces, we find a narrow region where the multivalent counterions are partially depleted from the vicinity of the surfaces because of the salt image repulsions. This depletion effect arises because of the inhomogeneous distribution of the monovalent salt ions in the system (as the slabs are assumed to be impermeable to these ions) but its overall effect is quite weak and, when the surfaces are randomly charged, gives way to the singular attraction of counterions by disordered surface  charges at small separations.

This salt image (or screening-induced) depletion is in contrast with the depletion due to dielectric images, which in dielectrically inhomogeneous systems always dominates in the vicinity of the dielectric surfaces and creates an interfacial zone of complete depletion, even when the bounding surfaces of the slabs are randomly charged. The reason being the repulsive ion-surface potential generated by the dielectric images that diverges at a dielectric surface (with $\Delta>0$) 
in a way that overcomes the disorder-induced attraction experienced by the multivalent counterions at small distances from the surfaces. The amount of multivalent counterions accumulated at some short distance away the dielectric surface is, however, still overall enhanced because of the disorder-induced attraction.

We have also analyzed in detail the consequences that result from the interplay between charge disorder, dielectric and salt images, and the SC effects on the effective interaction pressure between the slabs. The interaction pressure shows in general a highly non-monotonic behavior as a function of the separation distance between the inner surfaces of the slabs. At small to intermediate separations (e.g., typically within a few nanometers), the SC effects can be quite significant. They contribute an attractive component to the net interaction pressure that opposes the repulsive osmotic pressure due to monovalent salt ions and the repulsive interaction between the mean surface charges on the two slabs. In the absence of disorder, this SC attraction between the two like-charged surfaces becomes dominant when the bulk concentration of multivalent counterions is increased and/or when the salt screening parameter or the dielectric discontinuity parameter are  decreased. Nevertheless, when the inter-surface separation decreases,  the pressure at very small separations becomes  repulsive once again since counterions are completely depleted from the slit due to dielectric image repulsions. In the presence of disorder, counterions are attracted to the surface much more strongly than in the absence of disorder and a much larger number of counterions are sucked into the slab region from the bulk. 
As a result, the SC attraction mediated by multivalent counterions between the disordered charged surfaces 
becomes increasingly more enhanced, especially as the disorder coupling parameter and/or the bulk concentration of multivalent counterions are increased.
The repulsive regime at small separations, which could stabilize the surfaces in the absence of disorder, is thus squeezed down to zero due to extremely large attractive pressures acting on the slabs.

We have presented our results in a rescaled (dimensionless) form, in terms of a few dimensionless parameters such as the rescaled screening parameter and the electrostatic and disorder coupling parameters. These parameters, for any given set of values, can be mapped to a wide range of values for the actual parameters, namely, the counterion and salt bulk concentrations, mean surface charge density, counterion valency, solvent dielectric constant, temperature, etc.  A few typical examples of the physical parameter values that can correspond to some typical values of the rescaled parameters are shown in Table \ref{table}; other sets of physical parameter values are conceivable, e.g., by using divalent and trivalent counterions or other surface charge densities, etc. It should be noted that the range of values we have used for the disorder coupling parameter, i.e., $\chi\simeq 0-4$, corresponds to typically small degrees of charge disorder with disorder variances of around $g\simeq 0- 0.08~{\mathrm{nm}}^{-2}$ that can be achieved  by relatively small surface density of  impurity charges ($\lesssim 0.1~{\mathrm{nm}}^{-2}$) as compared with the mean number of surface charges (typically $\sigma\lesssim 1~{\mathrm{nm}}^{-2}$). In other words, the effects due to the coupling between surface charge randomness and the SC electrostatics due to mobile multivalent counterions can be quite significant even at very small degrees of the surface charge disorder.

The regime of applicability of the dressed multivalent-ion theory (which can be justified strictly only in the case of highly asymmetric Coulomb fluids  \cite{SCdressed1}) has been discussed in detail elsewhere  \cite{SCdressed1,SCdressed2,SCdressed3,perspective,leili1,leili2} by making extensive comparison with implicit- and explicit-ion simulations in non-disordered systems, where it is found to be in the experimentally  accessible parameter space. This includes the situations where  electrostatic interactions are strong enough so that the effects of  multivalent counterions next to an oppositely charged boundary is adequately included on the lowest order single-particle level  
\cite{Netz01,AndrePRL,AndreEPJE,hoda_review,perspective,Naji_PhysicaA,asim}. The higher-order effects of multi-particle interactions between counterions are assumed to be sufficiently weak for moderate to high salt concentrations \cite{hoda_review,SCdressed1,SCdressed2,SCdressed3,perspective}. In particular, the present approach is  expected to be applicable for relatively small (a few mM) bulk concentrations of the multivalent counterions, which is in fact the typical case in most  experiments (see, e.g., Refs. \cite{rau-1,rau-2,Bloom2,Pelta,Plum,Raspaud,Pelta2,Yoshikawa1,Yoshikawa2,Savithri1987,deFrutos2005,Siber}). We expect that the previously determined regimes of validity hold also for the present case with  randomly charged surfaces, although this remains to be determined more systematically through extensive computer simulations that are  still missing for the disordered systems.

Our results yield concrete predictions that can be tested against these future simulations. The fingerprints of disorder effects can show up in surface-force measurements as well (see, e.g., Refs. \cite{disorder-PRL,jcp2010,pre2011,epje2012,jcp2012,book,jcp2014,speake,kim2,kim3} and references therein for charge disorder effects  in the context of Casimir force measurements and Refs. \cite{Meyer,Meyer2,klein,klein1,klein2} for force measurement between surfactant-coated surfaces in aqueous media). One should, however, note that the precise statistical distribution of charge disorder in real systems can be sample and material dependent and it can vary depending also on the method of preparation. These points need to be addressed first if the theoretical predictions are to be compared against experiments.

The present study  is based on a primitive model of Coulomb fluids and makes a few simplifying assumptions that have been discussed elsewhere in detail (see, e.g., Refs. \cite{SCdressed1,SCdressed2,SCdressed3,perspective,jcp2014}). For instance, we have neglected the solvent structure (see, e.g., Refs. \cite{Israelachvili,holm,benyaakov,Burak_solvent,Burak_solvent2,Ben-Yaakov2011,Ben-Yaakov2011b}  and references therein), the polarizability  of mobile ions (see, e.g., Refs. \cite{Bikerman,Demery} and references therein),  specific surface ion-adsorption effects \cite{Forsman06}, and the size and internal structure of the counterions \cite{frydel_rev}. For multivalent counterions that can be modeled as spherical particles, their finite size can be incorporated in our approach in a straightforward manner as discussed in detail in Ref. \cite{jcp2014}. 
It is important to note that  most multivalent counterion possess an internal charge distribution that can introduce higher-order multipolar effects; these effects can be quite significant  especially for  multivalent counterions that have an extended structure (see, e.g.,  Refs. \cite{multipoles,perspective} and references therein), including rod-like polyamines such as the trivalent spermidine and tetravalent spermine that have chain lengths of up to 1-1.5~nm \cite{spermidine_spermine}. We plan to address this issues in future publications. Other cases that can be studied in the present context in the future include spatially correlated surface charge disorder \cite{jcp2010,jcp2012,preprint4} as well as surfaces with annealed (mobile) disordered charges \cite{netz-disorder,netz-disorder2,andelman-disorder,disorder-PRL,jcp2010,safran1,safran2,safran3,Olvera0}, surfaces with partially quenched or partially annealed charges  \cite{partial,Hribar}, and also charge regulating surfaces \cite{Regulation,Regulation2,Regulation3,Olvera,Olvera0,natasha}.

Another point to be noted is that, in systems containing added monovalent salt, our theoretical approach may become invalid when the mean electrostatic potential near the randomly charged surfaces becomes large, in which case the validity of the underlying DH approximation used for the monovalent ions can break down \cite{SCdressed2}. This can happen when the disorder strength is large and/or when the dielectric discontinuity parameter is small. Other cases that goes beyond the present framework include the situation where nonlinear charge renormalization and/or Bjerrum pairing effects become relevant (see, e.g., Refs. \cite{Kjellander07,Alexander,Bjerrumpairing1,Bjerrumpairing2,Bjerrumpairing3}). These latter effects, however, turn out to be negligible in the regime of parameters that is of interest here \cite{SCdressed1,SCdressed2,SCdressed3,perspective}.  These and other possible issues such as higher-order virial corrections \cite{Netz01,AndrePRL,AndreEPJE,hoda_review,perspective,Naji_PhysicaA,asim} or the intermediate coupling effects (see, e.g., Refs. \cite{hoda_review,Naji_PhysicaA,perspective,AndreEPJE,Burak04,Hatlo-Lue,Weeks2,Santangelo}), the discrete nature of monovalent salt ions \cite{SCdressed2}, ion-ion excluded-volume repulsions \cite{overcharge2,Lozada-Cassou_review,Messina-Holm,Kjellander,frydel}, etc, that may be relevant especially at intermediate electrostatic couplings and/or large multivalent counterion concentrations,  remain to be elucidated in future simulations.

\begin{acknowledgements}

A.N. acknowledges partial support from the Royal Society, the Royal Academy of Engineering, and the British Academy (UK). H.K.M. acknowledges  support  from the School of Physics, Institute for Research in Fundamental Sciences (IPM), where she stayed as a short-term visiting researcher during the completion of this work.  R.P. acknowledges support from the ARRS through Grants No. P1-0055 and J1-4297.  We thank  E. Mahgerefteh for useful comments and discussions.

\end{acknowledgements}


\begin{thebibliography}{99}
\bibitem{VDW}
This refers to the isotherms as opposed to van der Waals interactions to be considered later.
\end{thebibliography}
\end{document}